\documentclass[aps,pra,10pt,twocolumn,superscriptaddress,floatfix]{revtex4-2}

\usepackage{amsmath,amssymb,bm}
\usepackage{graphicx}
\usepackage[hidelinks]{hyperref}
\usepackage[capitalise]{cleveref}
\usepackage{braket}
\usepackage{booktabs} 
\begin{document}

\title{Two-Qubit Module Based on Phonon-Coupled Ge Hole-Spin Qubits:\\
A Conditional Design Study for Fabrication and Readout at 1--4 K}

\author{D.-M.~Mei}
\email[Corresponding author: ]{Dongming.Mei@usd.edu}
\affiliation{Department of Physics, University of South Dakota, Vermillion, SD 57069, USA}

\author{S. A. Panamaldeniya}
\affiliation{Department of Physics, University of South Dakota, Vermillion, SD 57069, USA}

\author{K.-M. Dong}
\affiliation{Department of Physics, University of South Dakota, Vermillion, SD 57069, USA}

\author{S. Bhattarai}
\affiliation{Department of Physics, University of South Dakota, Vermillion, SD 57069, USA}

\author{A. Prem}
\affiliation{Department of Physics, University of South Dakota, Vermillion, SD 57069, USA}

\date{November 18, 2026}

\begin{abstract}
We present a conditional design study for two gate-defined germanium (Ge)
hole-spin qubits coupled to a shared phononic-crystal (PnC) defect mode,
with a proposed characterization program at $1$--$4~\mathrm{K}$.
We specify a candidate Ge/SiGe membrane process, electrostatic controls,
RF charge readout, and tests that distinguish cavity-mediated exchange
from direct interdot interactions. A reproducible one-dimensional
effective-mass calculation quantifies localization and residual tunneling
for two specified dot designs: the revised reference separation is
$100~\mathrm{nm}$ with a $20~\mathrm{nm}$ longitudinal oscillator length,
while a $50~\mathrm{nm}$ layout is retained as a more strongly coupled
comparison. Conditional laminate calculations quantify strain sharing and
natural curvature for explicitly stated $200~\mathrm{nm}$ core-only references.
Independent qubit tuning requires a full-rank gate response and a finite
trajectory that preserves charge occupation, bounds direct exchange, and
retains both cavity couplings. These device-specific results are not
established by the present local control formulation.
We derive the leading transverse cavity interaction and quantify thermal
occupation, cavity-induced relaxation, and number-noise channels for
specified exchange sequences. Orbital-parameter sensitivity and these
thermal calculations constrain coherent
operation at $1$--$4~\mathrm{K}$. We also give a $0$--$10^\circ$
sidewall-geometry sweep and the specification for the necessary open-boundary
elastic simulation; a device-specific quality-factor sweep has not yet
been established. The work reports a conditional design and reproducible
reduced-order calculations, without experimental device data or a
validated fabrication yield or entangling-gate fidelity.
We additionally report two reproducible screening models: a five-strip
Poisson half-space calculation gives a finite compensated path for
illustrative electric-field-to-spin responses, and a one-dimensional
edge-damped defect-chain calculation supplies an illustrative
nonmonotonic wall-angle-dependent $Q$. Neither model establishes
full-device electrostatic feasibility or a converged three-dimensional
acoustic quality factor.
\end{abstract}

\maketitle

\section{Introduction}

High-purity germanium (Ge) integrated with phononic-crystal (PnC) cavities has
emerged as a promising platform for warm-operating quantum hardware, combining
strong spin--orbit coupling, mature materials control, and engineered phonon
environments. Over the past decade, strained Ge/SiGe heterostructures have enabled
gate-defined hole-spin qubits with long coherence times, fast all-electrical control,
and multi-qubit operation, establishing Ge as a leading CMOS-compatible qubit
material alongside Si and III--V systems.~\cite{watzinger2018germanium,scappucci2021germanium,hendrickx2020fast}
In parallel, quantum-acoustics experiments have shown that GHz phononic modes in
nanostructured resonators can reach the quantum regime and serve as coherent buses
or memories when coupled to microwave or solid-state
qubits.~\cite{chu2017quantumacoustics,arrangoiz2019resolving,bienfait2019phonon}
Together, these advances motivate architectures in which Ge hole spins interact with
engineered phononic bandgaps and defect modes to suppress decoherence and mediate
entangling interactions.

Building on this foundation, Mei \emph{et al.} proposed a phonon-coupled Ge
hole-spin architecture in which qubits are hosted in a strained Ge quantum well and
coupled to localized PnC defect modes in a suspended Ge
membrane.~\cite{mei2025qst} That work provides design benchmarks for electric-dipole spin resonance
(EDSR), phonon-modified relaxation, and cavity-mediated coupling. Its
single-phonon assumptions do not establish coherent operation at
$1$--$4~\mathrm{K}$. The present study treats this warmer range as a
characterization objective and explicitly includes the thermal-mode
constraints in Sec.~\ref{subsec:operating_window}.
In related work, the GeQuLEP concept explores Ge-based quantum phononic
spectroscopy as a route to detect sub-MeV dark matter using long-lived, high-$Q$
phonon excitations in high-purity Ge, further highlighting the utility of engineered
Ge phononics for ultra-low-threshold sensing.~\cite{mei2025pdu}
Taken together, these studies identify a common opportunity: leveraging Ge phononics
to realize warm-operating quantum devices that naturally interface with both quantum
information processing and quantum sensing.

Beyond single- and few-qubit demonstrators, Ge hole-spin qubits provide a natural
route toward scalable processors and networked architectures. Group-IV compatibility,
isotopic purification, and CMOS-friendly processing enable dense quantum-dot arrays
and multi-qubit modules with all-electrical control, as demonstrated in Si/Ge
platforms with fast, high-fidelity two-qubit
gates.~\cite{hendrickx2020fast,lawrie2020quantum}
Coupling Ge spin qubits to on-chip microwave or phononic resonators further supports
modular designs in which local registers are linked by bosonic buses or interfaced to
photonic channels for long-distance entanglement distribution and quantum
networking.~\cite{bienfait2019phonon,kimble2008quantum,wehner2018quantum}
A phonon-engineered Ge platform could connect local spin registers to
such buses. Whether useful memory and entangling operations survive at
$1$--$4~\mathrm{K}$ depends on the measured mode occupation, relaxation,
and dephasing rather than on bath temperature alone.

A further motivation for the present two-qubit design comes from the coherence
limit observed in isotopically purified Ge. A benchmark study of shallow donor
spins in isotopically enriched Ge at approximately $350$ mK and
$0.44$ T reported a relaxation-limited coherence time,
$T_2=2T_1=1.2~\mathrm{ms}$, implying that the pure-dephasing contribution
$T_{\phi}^{-1}$ was negligible within the resolution of that experiment.\cite{sigillito2015electron}
Although this donor-spin benchmark is not itself a gate-defined Ge hole-spin qubit result,
it provides an important materials reference point for the relaxation-limited regime.
Because the published value was reported as a relaxation-limited benchmark rather
than as a full error-budget decomposition of all residual dephasing channels, it
does not by itself provide a precise quantitative upper bound on
$T_{\phi}^{-1}$. This observation nevertheless identifies a central materials
principle: when the nuclear-spin bath is suppressed by isotopic enrichment and
residual impurity disorder is minimized by high-purity crystal growth, spin
coherence can approach the relaxation limit in favorable operating regimes. High
chemical purity can also reduce impurity-related local strain and charge
disorder; however, in a Ge/SiGe quantum-well device, residual strain disorder can
also arise from alloy disorder, interface roughness, misfit dislocations,
membrane release, and gate-induced stress. Recent demonstrations of
nuclear-spin-depleted $^{70}\mathrm{Ge}/^{28}\mathrm{Si}^{70}\mathrm{Ge}$ quantum wells, together with
the availability of spin-zero Ge isotopes such as $^{70}\mathrm{Ge}$ and $^{74}\mathrm{Ge}$,
therefore motivate a two-qubit experiment in high-purity, isotopically enriched
Ge to measure $T_2$, extract or bound $T_\phi$, and test whether suppression of
environmental dephasing is preserved after adding gates, a suspended membrane, a
PnC cavity, and interqubit coupling.\cite{moutanabbir2024nuclearspin,daoust2026nuclearspinfree}

Motivated by these advances, the present work is organized around a set of concrete,
experimentally testable research questions that go beyond proof-of-principle
single-qubit demonstrations. First, can PnC bandgaps engineered in a
suspended Ge membrane suppress phonon-induced relaxation and preserve spin lifetimes
$T_{1}$ at elevated cryogenic temperatures ($1$--$4~\mathrm{K}$)? Second, can a
single localized phononic defect mode provide sufficiently strong and coherent
spin--phonon coupling to mediate resolvable two-qubit interactions between spatially
separated Ge hole-spin qubits? Third, can high-fidelity readout and control be
achieved using radio-frequency charge sensing and cryogenic amplification without
reliance on dilution refrigeration? Addressing these questions will test whether
phonon-engineered Ge hole-spin devices can serve as a scalable, warm-operating
quantum hardware platform.

In this work we examine fabrication and control requirements for a
candidate two-qubit module that
can serve as a building block for larger Ge-based quantum processors and quantum
sensors. The module consists of two gate-defined hole-spin qubits formed in a
strained Ge quantum well, with a revised reference nominal separation of
$100~\mathrm{nm}$, and coupled
through a localized PnC defect mode in a suspended Ge membrane. We analyze the
heterostructure stack, electrostatic layout, and phononic design needed to
support strong spin--phonon coupling while maintaining high $g$-factor
tunability and robust charge stability. Particular attention is given to
fabrication choices compatible with established Ge growth and processing flows,
including etch chemistries, membrane thickness, sidewall geometry, membrane
stress, surface passivation, and metallization schemes. The results presented
here constitute a conditional design study and proposed experimental
program; a complete device specification and experimental realization
remain future work.

This work combines a proposed integration of PnC cavities with
gate-defined Ge hole-spin qubits, a candidate nanofabrication process flow, cryogenic
readout architecture, and conditional performance benchmarks at
$1$--$4~\mathrm{K}$. Performance values inherited from Ref.~\cite{mei2025qst} are modeling
inputs. New results here are reduced-order calculations of dot localization,
tunnel splitting, strain sharing, sidewall geometry, a five-strip
compensation trajectory, and a one-dimensional acoustic taper sweep,
together with explicit attribution tests. These calculations do not replace
self-consistent multiband electrostatics or an open-boundary elastic model
of the final patterned device.

\begin{table*}[t]
\caption{Scope of the present design study relative to Ref.~\cite{mei2025qst}.
The numerical examples added here are conditional reduced-order calculations;
device-specific fabrication and coherent-gate validation remain outstanding.}
\label{tab:novelty_ref7}
\begin{ruledtabular}
\begin{tabular}{p{0.25\textwidth}p{0.31\textwidth}p{0.36\textwidth}}
Aspect & Prior modeling inputs & Present manuscript \\
\hline
Qubit layout & Generic gate-defined Ge qubits & Defined oscillator length;
$50$ and $100~\mathrm{nm}$ double-well calculations, tunnel exchange,
and sensitivity to the prescribed confinement/barrier/charging inputs;
ideal strip-gate field spreading \\
Released membrane & Ge/PnC design benchmarks & Candidate release sequence;
conditional strain-sharing and curvature calculations \\
Etching tolerance & Nominal cavity parameters & Explicit taper geometry;
requirements for a converged $Q(\theta)$ calculation, which is not yet available \\
Control and attribution & Spin--phonon and dispersive-coupling estimates &
Five-control virtual-gate protocol; direct-exchange controls; proposed
paired $T_1$, $T_2^*$ and $T_2$ measurements \\
Thermal operation & Single-phonon regime & Finite-occupation relaxation and thermal phase-noise channel
calculations for specified exchange sequences \\
\end{tabular}
\end{ruledtabular}
\end{table*}
The remainder of this paper is organized as follows.
Section~\ref{sec:device_design} introduces the overall two-qubit device concept and
electrostatic layout, while Sec.~\ref{sec:heter} specifies the SiGe/Ge
heterostructure and materials stack that support the
phononic-crystal--integrated module. In Sec.~\ref{sec:fabrication} we outline a
compatible nanofabrication process flow, including mesa definition, gate patterning,
membrane release, and RF/DC wiring. Section~\ref{sec:readout} presents the
spin-to-charge conversion protocols, charge-sensor design, and RF signal chain for
high-fidelity qubit readout, and Sec.~\ref{sec:measurements} details the cryogenic
measurement environment for operation at $1$--$4~\mathrm{K}$.
In Sec.~\ref{sec:benchmarks} we propose an experimental program with target metrics
for single- and two-qubit control and for validating the engineered phononic
environment. Section~\ref{sec:discussion} discusses anticipated performance,
extensions of the architecture, and broader implications for scalable Ge-based
quantum processors and quantum sensors. Finally, Sec.~\ref{sec:conclusions}
summarizes the main design outcomes and highlights paths toward multi-qubit scaling
and integration into larger quantum-network architectures.
This organization is intended to make the scope of the paper explicit: the
manuscript is a device-level design study and validation roadmap, not an
experimental report of a realized two-qubit device. The distinction from
Ref.~\cite{mei2025qst} is summarized in Table~\ref{tab:novelty_ref7}.

\section{Device Concept and Two-Qubit Layout}
\label{sec:device_design}

\begin{figure}[t]
\centering
\includegraphics[width=\linewidth]{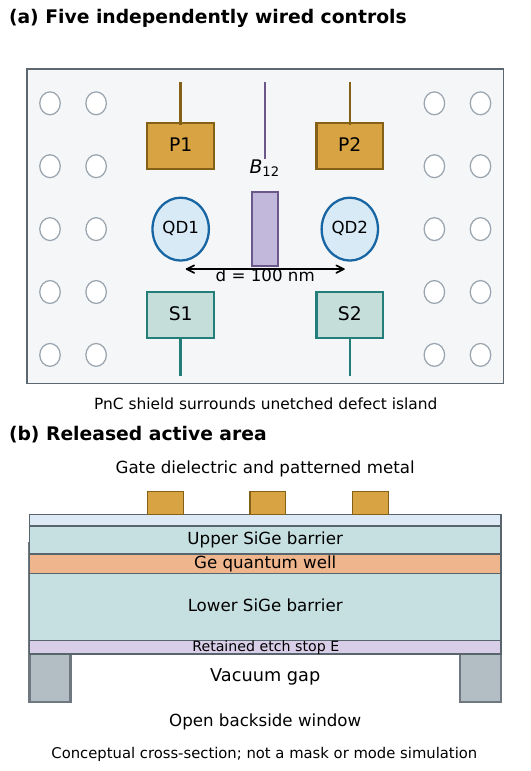}
\caption{Revised conceptual layout, not to scale. (a) Plungers $P_1,P_2$,
interdot barrier $B_{12}$, and independently wired local shape gates
$S_1,S_2$ provide five candidate controls for two charge potentials,
tunnel coupling, and two spin frequencies. Reservoir barriers and sensor
leads are omitted for clarity. (b) The Ge quantum well lies within the
released Ge/SiGe membrane; a vacuum gap separates the active area from
the substrate. The nominal reference dot separation is $100~\mathrm{nm}$. The PnC shield
surrounds an unetched dot-bearing defect island; its final mode and anchor
geometry require device-specific modeling.}
\label{fig:device_concept}
\end{figure}

The two-qubit module consists of a pair of gate-defined hole quantum dots,
QD$_1$ and QD$_2$, formed in a compressively strained Ge quantum well grown on a
Si substrate and separated from metallic top gates (plunger and barrier gates)
by a thin dielectric stack, as shown in Fig.~\ref{fig:device_concept}. Metallic
surface gates locally deplete the high-mobility two-dimensional hole gas
(2DHG). The revised reference layout uses a $100~\mathrm{nm}$ nominal
separation and separately wired plunger, barrier, and local shape gates.
The earlier $50~\mathrm{nm}$ spacing is examined as a compact comparison
in Sec.~\ref{subsec:gate_qubits}. The desired interaction-stage tunnel
coupling must be measured and suppressed; proximity alone does not
establish independent spin control.

Crucially, the Ge quantum well is integrated with an underlying
PnC cavity patterned into a suspended membrane region. The
PnC cavity acts as both a spectral and spatial filter for lattice vibrations:
it suppresses coupling to bulk substrate phonons while enhancing interaction
with a small number of localized phonon modes that spatially overlap the
double-dot region. As a result, phonons confined in the cavity can mediate
coherent spin--spin coupling between the two hole-spin qubits and provide
additional control parameters—through the phononic bandgap and cavity
linewidth—for tailoring dissipation, relaxation, and qubit--environment
coupling. Figure~\ref{fig:device_concept} summarizes the device concept and
introduces the geometric layout analyzed in the following subsections.

The nominal spin--phonon and cavity parameters are taken from the
modeling framework of Ref.~\cite{mei2025qst}. They are treated as inputs
to the design tests below, not as new finite-element calculations for the
revised membrane and double dot. In particular, changing dot confinement,
separation, or release-induced strain requires recalculating the
spin--phonon matrix elements.
\subsection{Gate-defined hole-spin qubits}
\label{subsec:gate_qubits}

A qubit is a single hole in a gate-defined strained Ge quantum well.
All-electrical EDSR uses spin--orbit coupling to drive spin rotations.
The compact geometry requires separate definitions of gate dimensions,
orbital extent, and interdot separation; charge and spin control will be
calibrated rather than assumed from the drawing.~\cite{hendrickx2020fast}

\paragraph{Orbital extent and revised reference spacing.}
We distinguish the lithographic gate dimensions, the separation $d$ between
potential minima, and the orbital confinement length. For a locally harmonic
longitudinal potential, we define
\begin{equation}
 \ell=\sqrt{\frac{\hbar}{m_\parallel\omega_x}},\qquad
 \sqrt{\langle(x-x_0)^2\rangle}=\frac{\ell}{\sqrt{2}}.
 \label{eq:orbital_length}
\end{equation}
The second relation applies to an isolated harmonic ground state.
The calculated localized-orbital rms widths for the reference design
are $13.72$--$14.19$ nm; individual values accompany the numerical data. A length of
$100$--$150~\mathrm{nm}$ cannot be used as the harmonic oscillator length of
two separately localized dots at $d=50~\mathrm{nm}$ without an additional
confinement mechanism. We therefore use $d=100~\mathrm{nm}$ and
$\ell=20~\mathrm{nm}$ as a revised reference geometry for an orbital
consistency calculation. The compact $d=50~\mathrm{nm}$ case is retained only
as a comparison, with a tighter assumed confinement of
$\ell=15~\mathrm{nm}$. Neither geometry is presented as a validated
electrostatic design.

\paragraph{Illustrative potential and wavefunctions.}
To make the localization and residual tunneling explicit, we solve
\begin{equation}
 H_x=-\frac{\hbar^2}{2m_\parallel}\frac{d^2}{dx^2}+V(x;A),
 \label{eq:orbital_hamiltonian}
\end{equation}
with
\begin{align}
 V(x;A)&=\frac{\hbar^2(|x|-d/2)^2}{2m_\parallel\ell^4}
       +A\,b(x),\label{eq:double_dot_potential}\\
 b(x)&=\begin{cases}
 \cos^2[\pi x/(2b_0)],& |x|<b_0,\\
 0,& |x|\geq b_0,
 \end{cases}\quad b_0=15~\mathrm{nm}.\nonumber
\end{align}
Here $m_\parallel=0.05m_0$ is an illustrative low-density Ge in-plane mass,
motivated by the extrapolated value in Ref.~\cite{lodari2019light}; it is not
an extracted mass for the proposed suspended device. The barrier parameter
$A=0$, $30$, or $60~\mathrm{meV}$ is an added potential energy, not an
experimentally calibrated gate voltage. Both $A$ and $b_0$ are
prescribed potential parameters. A physical gate changes the full
confinement potential and can shift both dot minima, their sizes, their
charge potentials, and the tunnel coupling. No independent voltage
control of $A$ is assumed. The reference local harmonic spacing is
$\hbar\omega_x=3.81~\mathrm{meV}$.

\begin{figure*}[t]
 \centering
 \includegraphics[width=0.96\textwidth]{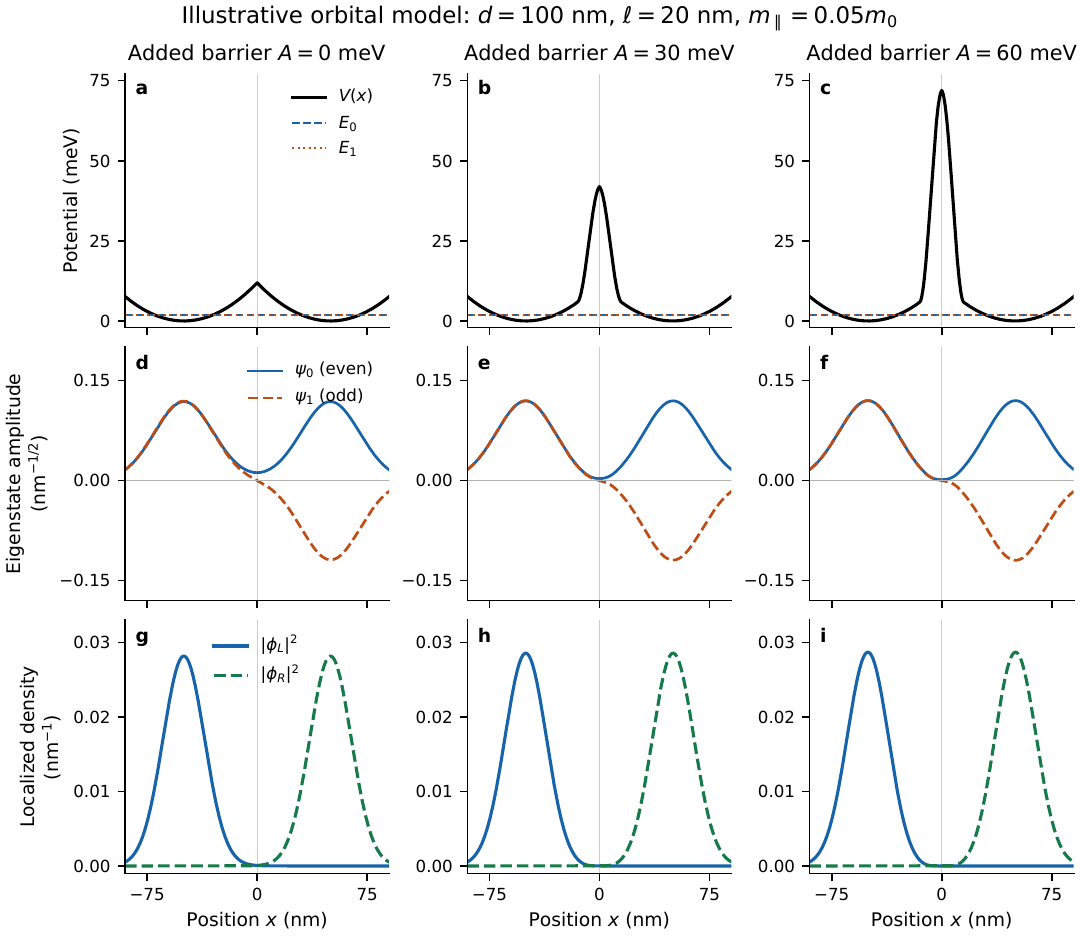}
 \caption{Illustrative longitudinal orbital calculation for the revised
 reference geometry, $d=100~\mathrm{nm}$ and $\ell=20~\mathrm{nm}$.
 Columns use added barrier energies $A=0$, $30$, and $60~\mathrm{meV}$.
 (a)--(c) Assumed potential and lowest two energies, which are nearly
 indistinguishable on this energy scale.
 (d)--(f) Normalized even and odd eigenfunctions.
 (g)--(i) Densities of the corresponding orthonormal localized orbitals.
 The barrier energies are model inputs; no gate-voltage calibration,
 charge-occupation validation, or multiband device simulation is implied.}
 \label{fig:dot_orbitals}
\end{figure*}

We use a second-order finite-difference grid with $0.1~\mathrm{nm}$ spacing
and zero-wavefunction boundaries at $x=\pm150~\mathrm{nm}$. Halving the grid
spacing changes the reported $t_c$ by less than $0.02\%$; a boundary check at
$\pm200~\mathrm{nm}$ is included with the calculation files.
Figure~\ref{fig:dot_orbitals} shows the potential and the two lowest
eigenfunctions at all three barrier settings. At zero detuning these
eigenfunctions are delocalized even and odd states, not one-hole states
assigned to individual dots. The orthonormal localized orbitals
$\phi_{L,R}$ are obtained by diagonalizing the position operator in this
two-state subspace; equivalently, up to phases,
$\phi_{L,R}=(\psi_0\pm\psi_1)/\sqrt{2}$. Their inner product vanishes by
construction. To quantify spatial leakage without confusing it with that
inner product, we report the wrong-half probability
$P_{\rm wrong}=\int_0^\infty |\phi_L(x)|^2dx$ and
$t_c=(E_1-E_0)/2$ in Table~\ref{tab:orbital_check}.

\begin{table*}[t]
 \caption{Numerical orbital consistency check. $A$ is the added central
 barrier energy, not the total barrier height. The last column uses the
 illustrative Hubbard estimate $J_{\rm ex}\simeq4t_c^2/U_{\rm eff}$ with
 $U_{\rm eff}=5~\mathrm{meV}$ assumed. It does not include direct Coulomb
 exchange, spin--orbit anisotropy, or excited-orbital corrections. The
 calculation does not establish that these potentials are accessible to the
 proposed gates or that the device remains in $(1,1)$.}
 \label{tab:orbital_check}
 \begin{ruledtabular}
 \begin{tabular}{rrrrrr}
 $d$ (nm) & $\ell$ (nm) & $A$ (meV) & $t_c/h$ (GHz)
 & $P_{\rm wrong}$ & $J_{\rm ex}/h$ (MHz) \\
 100 & 20 & 0  & 2.537   & $4.09\times10^{-4}$ & 21.30 \\
 100 & 20 & 30 & 0.2944  & $1.09\times10^{-5}$ & 0.2867 \\
 100 & 20 & 60 & 0.07102 & $1.31\times10^{-6}$ & 0.01669 \\
 50  & 15 & 0  & 96.72   & $2.01\times10^{-2}$ & $3.10\times10^4$ \\
 50  & 15 & 30 & 8.889   & $4.05\times10^{-4}$ & 261.4 \\
 50  & 15 & 60 & 1.927   & $4.01\times10^{-5}$ & 12.29 \\
 \end{tabular}
 \end{ruledtabular}
\end{table*}

The comparison shows why localization alone is insufficient for isolating
the phonon bus. For $d=50~\mathrm{nm}$, even the largest illustrated barrier
leaves $t_c/h=1.93~\mathrm{GHz}$ and an estimated
$J_{\rm ex}/h=12.3~\mathrm{MHz}$ for the assumed
$U_{\rm eff}=5~\mathrm{meV}$. For the $100~\mathrm{nm}$ reference and
$A=60~\mathrm{meV}$, the corresponding values are $71.0~\mathrm{MHz}$ and
$16.7~\mathrm{kHz}$. Writing this exchange energy as $J_H=J_{\rm ex}$, its flip--flop
coefficient is $J_d=J_H/(2\hbar)$. Thus $J_d/2\pi=8.34$ kHz in the
reference example, below a chosen attribution margin of $25$ kHz
(one tenth of $J_c/2\pi=0.25$ MHz). This comparison uses the
flip--flop coefficient, rather than the full singlet--triplet exchange
energy, consistently throughout the sensitivity study below. The two
examples change both separation and confinement length, so they
are a comparison of two assumed designs, not a spacing-only sweep. This conditional
example motivates a less compact layout; it does not demonstrate that the
fabricated device satisfies the bound. The source package retains the
continuous barrier scans for both assumed geometries.

\paragraph{Sensitivity of the conditional exchange bound.}
The reference point is not a robustness result. We therefore vary one
prescribed input at a time in the same Hamiltonian: the charge-transfer
denominator $U_{\rm eff}=1$--$10$ meV, barrier half-support
$b_0=10$--$20$ nm, added barrier energy $A=30$--$90$ meV, and
oscillator length $\ell=16$--$26$ nm. The other inputs retain their
reference values $d=100$ nm, $\ell=20$ nm, $b_0=15$ nm,
$A=60$ meV, and $U_{\rm eff}=5$ meV. These ranges are declared
sensitivity sweeps, not measured uncertainty intervals or gate-voltage
ranges. In particular, changing $U_{\rm eff}$ changes the exchange
estimate but does not change the one-particle eigenproblem.

\begin{figure*}[t]
 \centering
 \includegraphics[width=0.96\textwidth]{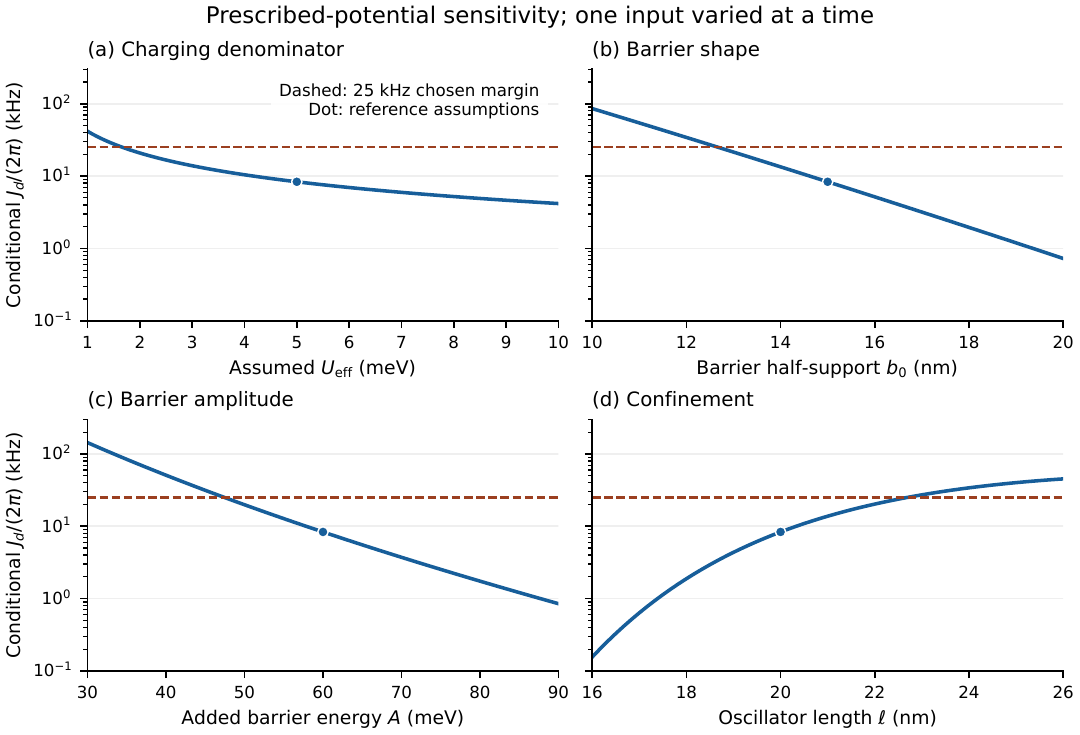}
 \caption{Sensitivity within the prescribed one-dimensional potential.
 Each panel varies one input, holding the others at
 $d=100$ nm, $\ell=20$ nm, $b_0=15$ nm, $A=60$ meV and
 $U_{\rm eff}=5$ meV. Points mark the common reference assumptions.
 All panels show the direct flip--flop coefficient
 $J_d/(2\pi)=2t_c^2/(hU_{\rm eff})$, with the same chosen
 $25$ kHz attribution margin (dashed). The sweeps do not describe
 correlated fabrication errors or a calibrated gate trajectory and do
 not establish the attainable operating region of a physical device.}
 \label{fig:dot_tunnel}
 \label{fig:dot_sensitivity}
\end{figure*}

Figure~\ref{fig:dot_tunnel} and Table~\ref{tab:orbital_sensitivity}
show that the chosen $J_d/(2\pi)<25$ kHz margin can be crossed by
changing assumptions that the present electrostatic model does not fix.
For example, decreasing $b_0$ from $15$ to $12$ nm increases
$J_d/(2\pi)$ from $8.34$ to $34.24$ kHz, while $A=45$ meV or
$\ell=24$ nm gives $31.41$ or $33.97$ kHz, respectively.
At the reference potential alone, the same approximate bound requires
$U_{\rm eff}>1.67$ meV. This is a conditional algebraic requirement,
not a predicted charging energy or proof of charge stability.
The scan endpoints were checked at $0.05$ nm grid spacing and a
$\pm200$ nm domain; the largest changes in $t_c$ were below
$0.02\%$ for grid refinement and $0.001\%$ for domain extension.
Numerical convergence therefore resolves the stated model, while
electrostatic uncertainty still limits its device interpretation.

\begin{table}[t]
 \caption{Selected conditional sensitivity results for $d=100$ nm.
 Only the listed input differs from the reference
 $\ell=20$ nm, $b_0=15$ nm, $A=60$ meV,
 $U_{\rm eff}=5$ meV. Charging energies and barrier shapes are
 assumptions, not extracted device parameters.}
 \label{tab:orbital_sensitivity}
 \begin{ruledtabular}
 \begin{tabular}{lrr}
 Changed input & \shortstack{$t_c/h$\\(MHz)} &
 \shortstack{$J_d/(2\pi)$\\(kHz)} \\
 Reference & 71.02 & 8.343 \\
 $U_{\rm eff}=1$ meV & 71.02 & 41.71 \\
 $U_{\rm eff}=10$ meV & 71.02 & 4.171 \\
 $b_0=12$ nm & 143.9 & 34.24 \\
 $b_0=18$ nm & 34.37 & 1.954 \\
 $A=45$ meV & 137.8 & 31.41 \\
 $\ell=24$ nm & 143.3 & 33.97 \\
 \end{tabular}
 \end{ruledtabular}
\end{table}

\paragraph{Requirements for a device electrostatic solution.}
The localized density centroids are not fixed exactly at the potential
minima: their separation changes from $99.92$ to $101.36$ nm across
the three reference barrier settings. The specified semiconductor stack
places the well center about $40$ nm below its top surface, before
including the gate dielectric. A $b_0=15$ nm model barrier has only
$30$ nm total support and $15$ nm full width at half maximum. Its
attainability cannot be inferred from the drawn gate linewidths:
electrostatic spreading, screening, and compensation by adjacent gates
must be calculated.

The analytical strip-gate diagnostic in
Appendix~\ref{app:strip_gate_diagnostic} quantifies lateral field
spreading in a separate idealized boundary-value problem. It shows why
gate linewidth and fixed-point potential compensation alone do not
establish the prescribed barrier or preserve the orbitals.

A device calculation needs the complete electrode footprints and
connections, dielectric thicknesses and permittivities, fixed-charge and
reservoir boundary conditions, bias limits, and the released well profile
and strain. These inputs are not fully specified by the schematic.
The resulting three-dimensional gate potential must determine the
orbital confinement, dot motion, tunnel barrier, and charge-addition
spectrum along the same compensated path. An isolated barrier lever arm
does not establish that path. The required control and occupation tests
are stated in Sec.~\ref{subsec:charge_tuning}.

Realistic Ge hole calculations show that light-hole admixture and
excited states can affect the spin response~\cite{wang2024modeling}.
The scalar longitudinal calculation does not determine the $g$ tensor,
spin--phonon matrix elements, charging energy, or one-hole occupation.
The $g$ tensors and both spin--phonon couplings must therefore be
recomputed for the changed confinement, spacing, and released strain,
using the same wavefunctions and strain mode used in the gate analysis.
The MHz coupling inputs used elsewhere remain conditional until that
joint calculation or device calibration is available.
\subsection{PnC cavity and phonon bus}
\label{subsec:pncbus}

The proposed bus is a localized mode of a suspended Ge/SiGe PnC membrane.
For the analytical geometry references, the uncoated semiconductor core
has thickness $t_0=200~\mathrm{nm}$ and circular holes on a triangular
lattice with nominal pitch $a=1.0~\mu\mathrm{m}$ and top radius
$r_t=0.30a$. The candidate release route additionally retains an
etch-stop layer $E$ beneath this core; its properties and the finite
defect, shield, support, gate, and coating geometry remain to be specified.
The actual suspended stack is therefore not defined by $t_0$ alone.
The dot-bearing defect remains unetched, whereas the PnC holes must
penetrate the retained $E$ layer. These reference dimensions do not
establish a complete 4--8 GHz bandgap or a localized 5--6 GHz mode.
Those properties require a band-structure and finite-device calculation
for the selected full stack and boundary conditions.

Dots must overlap the relevant components of the mode's strain tensor.
A displacement antinode alone need not maximize strain. In a multiband
description the coupling is
\begin{equation}
g_i=\frac{1}{\hbar}\left|\left\langle\uparrow_i\left|
H_{\rm BP}[\varepsilon^{\rm zpf}(\mathbf r)]\right|\downarrow_i\right\rangle\right|,
\label{eq:strain_overlap}
\end{equation}
where $H_{\rm BP}$ is the Bir--Pikus strain Hamiltonian and the spinors
include spin--orbit mixing.~\cite{maier2013relaxation,mei2025qst}
The zero-point strain normalization, local dot envelope, and field
orientation all matter; etching errors can change $g_1$ and $g_2$
differently.

The values $g_i/2\pi$ of order MHz, $Q_m\sim1.8\times10^4$, and
$T_1\sim1~\mathrm{ms}$ are prior-model benchmarks. They have not been
established for the revised device or simultaneously at a warm, coupled
operating point. We therefore use them only in explicitly conditional
estimates and retain independent measurements of $g_1$, $g_2$, $Q_m$,
mode occupation, and spin coherence as acceptance tests.

For the scalar estimates below, local spin phases are chosen so the
mode couplings are real; the displayed direct-interaction coefficient
is also taken real. A general complex or anisotropic direct interaction
requires the corresponding Hermitian-conjugate terms and separate
calibration. Within this convention, a minimal exchange model retaining a residual longitudinal term is
\begin{equation}
 \begin{aligned}
 \frac{H}{\hbar}&=\omega_c b^\dagger b+
 \sum_{i=1}^{2}\frac{\omega_i}{2}\sigma_i^z\\
 &\quad+\sum_{i=1}^{2}g_i(\sigma_i^+b+\sigma_i^-b^\dagger)\\
 &\quad+J_d(\sigma_1^+\sigma_2^-+\sigma_1^-\sigma_2^+)\\
&\quad+\zeta_d\sigma_1^z\sigma_2^z.
 \end{aligned}
 \label{eq:two_spin_bus}
\end{equation}
where $g_1$ and $g_2$ may differ after fabrication and $J_d$ denotes the
residual direct flip--flop coupling; $\zeta_d$ is the direct longitudinal coefficient. All rates in this equation are angular
frequencies; quoted MHz values refer to division by $2\pi$.
\paragraph{Residual longitudinal interaction.}
The residual longitudinal coefficient $\zeta_d$ is calibrated separately. For the
spin-conserving Hubbard estimate used in Sec.~\ref{subsec:gate_qubits}, the complete direct spin
Hamiltonian, up to a constant, is
\begin{equation}
 \frac{H_H}{\hbar}=\frac{J_H}{2\hbar}
 (\sigma_1^+\sigma_2^-+\sigma_1^-\sigma_2^+)
 +\frac{J_H}{4\hbar}\sigma_1^z\sigma_2^z.
 \label{eq:direct_hubbard_full}
\end{equation}
Thus $J_d=J_H/(2\hbar)$ and the longitudinal angular coefficient is
$\zeta_d=J_H/(4\hbar)=J_d/2$ in this restricted model. The longitudinal
term is constant in the single-excitation subspace, but contributes
conditional phases for a full two-qubit gate. General hole-spin
anisotropy can change these relations. Conditional spectroscopy must
therefore bound the longitudinal and transverse direct terms separately;
the longitudinal term is restored in any full gate calculation. The
second-order cavity interaction derived here remains transverse exchange.

For $\Delta_i=\omega_i-\omega_c$ with
$|\Delta_i|\gg |g_i|\sqrt{\bar n+1},\kappa$, elimination of the mode to
second order gives
\begin{align}
 \frac{H_{\rm disp}}{\hbar}&=\omega_c b^\dagger b+
 \sum_i\left[\frac{\omega_i+\chi_i}{2}
       +\chi_i b^\dagger b\right]\sigma_i^z\nonumber\\
 &\quad+(J_d+J_c)(\sigma_1^+\sigma_2^-+\sigma_1^-\sigma_2^+)\nonumber\\
&\quad+\zeta_d\sigma_1^z\sigma_2^z,\\
 \chi_i&=\frac{g_i^2}{\Delta_i},\qquad
 J_c=\frac{g_1g_2}{2}\left(\frac{1}{\Delta_1}+
                                      \frac{1}{\Delta_2}\right).
 \label{eq:dispersive_exchange}
\end{align}
This leading cavity-mediated contribution is transverse exchange, not a
$ZZ$ interaction.
It supports exchange spectroscopy and, when the qubits are mutually
resonant, iSWAP-type operations. A controlled-phase protocol requires a
separately derived higher-order or driven interaction and is not inferred
from $J_c$ alone. The structure of Eq.~\eqref{eq:dispersive_exchange} is the
same as the dispersive cavity-bus interaction demonstrated in
Ref.~\cite{majer2007bus}.
\paragraph{Sidewall taper: geometry and the outstanding radiation-loss calculation.}
The wall angle $\theta$ is measured from the membrane normal. For a
through-hole traversing a retained-stack depth $h$, with fixed top radius $r_t$ and
inward taper,
\begin{equation}
r_b=r_t-h\tan\theta,\qquad
r_{\rm vol}=\left[\frac{r_t^2+r_t r_b+r_b^2}{3}\right]^{1/2},
\label{eq:taper_geometry}
\end{equation}
where $r_{\rm vol}$ is the radius of a cylinder with the same removed
volume. The depth $h$ includes the retained etch-stop layer $E$ and
every retained layer penetrated by the hole; sacrificial buffer removed
during release is excluded. In particular, $h\geq200~\mathrm{nm}+t_E$
for the candidate semiconductor stack before any hole-wall coating.
Coating thickness and the measured final profile must be included separately.

Table~\ref{tab:taper_geometry} retains an exact geometric reference for
$h=200~\mathrm{nm}$ and $r_t=300~\mathrm{nm}$. It describes the core-only
reference, not the complete candidate membrane. At $10^\circ$ the radius
difference is 35.27 nm and the volume-equivalent radius changes by
$-5.82\%$. The corresponding depth correction is
$(h-200~\mathrm{nm})\tan10^\circ$; an additional 10 nm of etched depth, for
example, adds 1.76 nm to the radius difference. This is a geometric
sensitivity, not an assumed etch-stop thickness. At fixed mid-plane
radius, $r_{\rm vol}^2=r_{\rm mid}^2+h^2\tan^2\theta/12$ instead.

\begin{table}[t]
\centering
\caption{Core-only geometric reference for inward taper at fixed
$r_t=300~\mathrm{nm}$ and $h=200~\mathrm{nm}$. The actual etched depth
includes the retained $E$ layer. No entry is an acoustic simulation or a
quality factor.}
\label{tab:taper_geometry}
\begin{tabular}{rccc}
\hline
$\theta$ (deg) & $r_t-r_b$ (nm) & $r_b$ (nm) & $r_{\rm vol}$ (nm)\\
\hline
0 & 0.00 & 300.00 & 300.00\\
2 & 6.98 & 293.02 & 296.51\\
4 & 13.99 & 286.01 & 293.04\\
6 & 21.02 & 278.98 & 289.55\\
8 & 28.11 & 271.89 & 286.06\\
10 & 35.27 & 264.73 & 282.55\\
\hline
\end{tabular}
\end{table}

A tapered lossless structure can radiate elastic energy through its
supports even though taper is not itself a material-damping mechanism.
Its radiation-limited quality factor can be calculated without measured
intrinsic-loss data, provided the finite defect, shield, full stack, and
support geometry are specified. With open radiation boundaries or
perfectly matched layers (PMLs), a complex eigenfrequency gives
\begin{equation}
Q_{\rm rad}=\frac{\operatorname{Re}\tilde\omega}
{2|\operatorname{Im}\tilde\omega|},\qquad
Q_{\rm tot}^{-1}=Q_{\rm rad}^{-1}+Q_{\rm int}^{-1}+Q_{\rm load}^{-1}.
\label{eq:taper_q}
\end{equation}
Unknown intrinsic and loading losses limit a prediction of $Q_{\rm tot}$;
they do not prevent computing $Q_{\rm rad}$ for a specified device.

The finite defect dimensions and hole modifications, shield-cell count
and termination, supports, retained $E$ layer, and gate/coating geometry
remain unspecified. Selecting and documenting these features is
remaining design work required to define the proposed resonator, rather
than an intrinsic obstacle to simulating it. No full three-dimensional device $Q(0)$ or
$Q(\theta)$ curve has been established for this candidate. A complete
geometry must first support a localized baseline mode in the intended
frequency range. The same geometry and materials must then enter the
static release/cooling calculation and the prestressed open-boundary
mode calculation at $\theta=0,2,4,6,8,10^\circ$. Mode tracking and
mesh/PML/substrate convergence must distinguish a physical change in
leakage from a change in mode identity or numerical error. The common
validation requirements are summarized in
Appendix~\ref{app:common_validation}. The calculation must report
$f(\theta)$, $Q_{\rm rad}(\theta)$ and localization, including
nonmonotonic behavior or loss of the localized mode if encountered.
The geometric radius sweep is not that result. No numerical fabrication taper tolerance for the proposed device
is established here.

\paragraph{Exploratory 1D defect-chain eigenvalue calculation (not device FEM).}
To test the assumption that wall taper must monotonically lower $Q$,
we additionally solved a specified \emph{one-dimensional reduced acoustic
surrogate} with 21 mass sites, alternating ligament springs
($k_A:k_B=1:2.2$), a central mass defect (factor 1.55), and terminal
viscous dashpots ($c=0.12$ in normalized units). The PnC and defect
hole top radii are $300$ and $230$ nm, respectively, with lattice pitch
$1~\mu$m and a $200$ nm core-only etch depth. At angle $\theta$,
the effective radius is $r_{\rm eq}^2=(r_t^2+r_tr_b+r_b^2)/3$;
we scale site mass with solid area fraction
$1-\pi r_{\rm eq}^2/a^2$ and ligament stiffness with
$(a-2r_{\rm eq})^3$. The central mode is selected by its five-site
defect localization and followed across angles. The absolute
frequency unit is \emph{chosen} to set the $0^\circ$ mode at 6 GHz,
not determined from Ge elastic constants.
\begin{table}[t]
\caption{One-dimensional damped-chain \emph{surrogate} output, not
$Q_{\rm rad}$ of the proposed 3D Ge/SiGe cavity. Frequency is
normalized to 6 GHz at zero taper; the five-site localization ratio is
$\sum_{|j-j_0|\le2}|u_j|^2/\sum_j|u_j|^2$.}
\label{tab:surrogate_q}
\begin{ruledtabular}\begin{tabular}{rccc}
$\theta$ (deg)&$f$ (GHz)&$Q_{\rm 1D}$&Localization\\
0 & 6.000 & 44,824 & 0.833 \\
2 & 6.107 & 45,144 & 0.833 \\
4 & 6.215 & 45,346 & 0.832 \\
6 & 6.323 & 45,442 & 0.832 \\
8 & 6.432 & 45,440 & 0.831 \\
10 & 6.542 & 45,348 & 0.830 \\
\end{tabular}\end{ruledtabular}
\end{table}
\begin{figure}[t]
\includegraphics[width=\columnwidth]{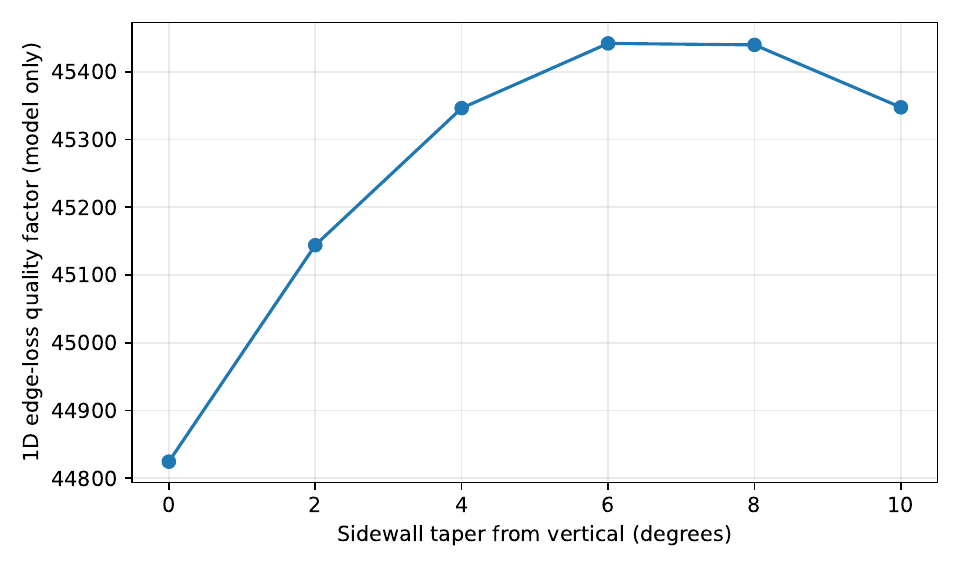}
\caption{Numerical eigenvalue sweep of the specified \emph{one-dimensional
edge-damped surrogate}; these are not full-stack elastic FEM predictions.
The increase and weak nonmonotonicity of $Q$ show why the sign of a
taper effect cannot be presumed.}
\label{fig:taper_chain_q}
\end{figure}
For this 21-site choice the computed $Q_{\rm 1D}$ changes from
$4.482\times10^4$ at $0^\circ$ to $4.535\times10^4$ at $10^\circ$.
Changing only the finite shield length gives very different
zero-angle values ($Q=9.33\times10^3$ with 17 sites and
$2.16\times10^5$ with 25 sites), so these values are \emph{not}
converged design predictions. Real anisotropic elasticity, the actual
finite defect/shield geometry, the retained etch-stop layer,
gates, prestress, substrate radiation and 3D absorbing boundaries are
absent. A validated $Q_{\rm rad}(\theta)$ or allowable fabrication
taper for the proposed device therefore remains outstanding.

\subsection{Fabrication tolerance and joint frequency recovery}
\label{subsec:sensitivity_analysis}

For $a=1.0~\mu\mathrm{m}$ and $r=0.30a$, a $5\%$ radius variation is
15 nm; across $r/a=0.2$--$0.4$ it is 10--20 nm. The
$\pm150~\mathrm{MHz}$ offset is retained solely as an illustrative
retuning challenge. It is not a simulated radius sensitivity or a
release-induced shift of the candidate stack. The prior-model coupling
and quality-factor values do not establish robustness against retained
layers, support leakage, coating stress, taper, or dot relocation.

A common magnetic-field shift cannot generally correct two unequal
qubits independently. Section~\ref{subsec:charge_tuning} therefore gives
a cross-compensated calibration procedure and acceptance tests for
occupation and both mode couplings. That procedure does not demonstrate
an accessible tuning trajectory, or imply that gate tuning repairs a
fabrication-induced loss of strain overlap. A finite-device acoustic
model and calibrated electrostatics remain needed to evaluate that
feasibility quantitatively.
\subsection{Operating window at 1--4 K}
\label{subsec:operating_window}

The proposed $1$--$4~\mathrm{K}$ range is a characterization objective.
At a fixed field orientation, the measured tensor for each dot sets
$f_i=\mu_B|\mathbf g_i\mathbf B|/h$; no common scalar $g$ or field
interval guarantees simultaneous resonance. For example, $f_i=6$ GHz
and $g_{\rm eff}=2$ require $B=0.214$ T. Local gate controls provide the
remaining differential tuning, subject to the charge and coupling tests
in Sec.~\ref{subsec:charge_tuning}.

At equal detuning $\Delta/2\pi=100~\mathrm{MHz}$ and
$g_1/2\pi=g_2/2\pi=5$--$10~\mathrm{MHz}$,
$J_c/2\pi=0.25$--$1~\mathrm{MHz}$ is an algebraic dispersive estimate.
The corresponding ideal exchange entangler has
$t_{\sqrt{\mathrm{iSWAP}}}=\pi/(4|J_c|)=0.125$--$0.5~\mu\mathrm{s}$,
and a full iSWAP takes twice as long. These times do not establish a gate
fidelity or a usable operating point. Both the dispersive approximation
and the coherence budget must be checked at the realized mode occupation.

The equilibrium occupation of a mode at temperature $T_m$ is
\begin{equation}
 \bar n=\left[\exp\left(\frac{\hbar\omega_c}{k_B T_m}\right)-1\right]^{-1}.
 \label{eq:thermal_mode_occupation}
\end{equation}
For $\omega_c/2\pi=6~\mathrm{GHz}$, $\bar n=3.00$ at $1~\mathrm{K}$
and $13.40$ at $4~\mathrm{K}$. A phononic bandgap suppresses continuum
coupling but does not by itself cool its defect mode. Taking $\kappa$ as the
mode energy-decay rate ($Q_m=\omega_c/\kappa$), a weak-dispersive estimate
of the cavity-induced population-relaxation rate is
\begin{equation}
 \Gamma_{1,i}^{(c)}\simeq
 (2\bar n+1)\frac{g_i^2\kappa}{\Delta_i^2+(\kappa/2)^2}.
 \label{eq:thermal_purcell}
\end{equation}
For the illustrative values $Q_m=1.8\times10^4$,
$g_i/2\pi=5~\mathrm{MHz}$ and $\Delta_i/2\pi=100~\mathrm{MHz}$,
Eq.~\eqref{eq:thermal_purcell} gives $[\Gamma_{1,i}^{(c)}]^{-1}$ of
$27.3~\mu\mathrm{s}$ at $1~\mathrm{K}$ and $6.87~\mu\mathrm{s}$ at
$4~\mathrm{K}$. These are conditional reduced-model estimates, not device
predictions, and other loss mechanisms shorten the total $T_1$ further.
In particular, the millisecond continuum-protected $T_1$ target cannot be
assigned unchanged to a strongly coupled or near-resonant operating point.
The $\chi_i b^\dagger b\sigma_i^z$ terms also convert occupation
fluctuations into phase noise.~\cite{gambetta2006dephasing,sears2012dephasing}

The dimensionless dispersive parameter
$\epsilon_i=|g_i|\sqrt{\bar n+1}/|\Delta_i|$ is approximately
$0.100$ and $0.190$ at $1$ and $4$ K, respectively, for
$g_i/2\pi=5$ MHz and $\Delta_i/2\pi=100$ MHz. It rises to $0.379$
at $4$ K for $g_i/2\pi=10$ MHz. In the latter case,
$n_{\rm crit}=\Delta_i^2/(4g_i^2)=25$ and about $16.5\%$ of the
thermal occupation distribution lies at $n\geq n_{\rm crit}$.
The $10$ MHz endpoint is consequently a parameter stress test, not a
quantitatively reliable dispersive gate operating point. It must be
treated with the full coupled model; the algebraic $0.125~\mu$s duration
does not establish an implementable gate. The $5$ MHz examples are also
conditional perturbative calculations, with their approximation error
requiring a full-model check before predicting device performance.

\paragraph{Thermal phase-noise diagnostic.}
Population relaxation alone does not assess the proposed warm operating
point. For $g_i/2\pi=5$ MHz and $\Delta_i/2\pi=100$ MHz,
$\chi_i/2\pi=0.25$ MHz. The exact long-time thermal-dephasing expression
within the dispersive oscillator model\cite{clerk2007thermal} gives inverse
rates of approximately $0.445~\mu$s at $1$ K and $0.177~\mu$s at $4$ K
for $f_c=6$ GHz and $Q_m=18000$. These are unrefocused asymptotic
model timescales, not Hahn-echo times or measured coherence.

We evaluate a specified ideal exchange pulse of duration
$\pi/(4|J_c|)$ and the same pulse with an ideal simultaneous midpoint
$X_1X_2$ echo, including its known final pulse frame. A number-diagonal
thermal Lindblad calculation retains all two-spin coherences and isolates
Stark phase noise. After deterministic local-$Z$ calibration, the
full-channel average fidelities are $0.585/0.659$ for plain/echoed
pulses at $1$ K and $0.508/0.513$ at $4$ K for equal couplings.
The ideal Bell-state diagnostic from $|10\rangle$ is nevertheless unity
in this restricted model: equal Stark shifts cancel in the
single-excitation subspace but not for arbitrary two-qubit states.

For $g_2/g_1=0.9$, we solve the differential bare-frequency compensation
while keeping the mean bare detuning at $100$ MHz. Thermal fluctuations
then cause an rms differential frequency of $0.334$ MHz at $1$ K and
$1.408$ MHz at $4$ K, compared with $2J_c/2\pi\simeq0.450$ MHz.
The full-channel diagnostics fall to $0.552/0.640$ and $0.376/0.427$,
respectively. Appendix~\ref{app:thermal_channel} specifies the channel, sequences,
and numerical checks; the accompanying files include the complete channels. These diagnostics omit
Purcell transitions, intrinsic spin noise, direct coupling, pulse errors,
and higher-order dispersive corrections. They are not expected device
fidelities or a general bound on warm operation. They do establish that
neither this passive sequence nor its single midpoint echo justifies a
$90$--$95\%$ gate-performance expectation. Such a target requires a
separately specified protocol and complete validation, potentially using
cooling, additional control, encoding, or a different operating point.

Near resonance, or when $|g_i|\sqrt{\bar n+1}/|\Delta_i|$ is not small,
the coupled thermal master equation must replace these asymptotic
expressions. Its mechanical terms are
$\kappa(\bar n+1)\mathcal D[b]\rho+
\kappa\bar n\mathcal D[b^\dagger]\rho$, where
$\mathcal D[L]\rho=L\rho L^\dagger-\{L^\dagger L,\rho\}/2$,
together with background qubit relaxation, excitation, and dephasing
terms. Cavity-induced losses already represented by this explicit mode
must not be added again as independent spin rates. No numerical
entangling-gate fidelity is assigned to the proposed device. A useful
gate would require a specified control or cooling strategy and a full
model checked against measured parameters. The calculations in
Appendix~\ref{app:thermal_channel} test two particular ideal sequences;
they do not rule out other operating points or protocols.

At $6$ GHz the Zeeman energy is only $24.8~\mu\mathrm{eV}$, whereas
$k_BT=86.2$--$344.7~\mu\mathrm{eV}$ at $1$--$4$ K. Ordinary
energy-selective loading is consequently thermally broadened.
Measurement-based initialization and a suitably resolved spin-to-charge
conversion window must be demonstrated; RF charge sensitivity alone does
not establish spin readout fidelity. A lower-temperature baseline is
appropriate if these conditions are not met.
\section{Heterostructure and Materials Stack}
\label{sec:heter}

\begin{figure*}[t]
\centering
\includegraphics[width=0.95\textwidth]{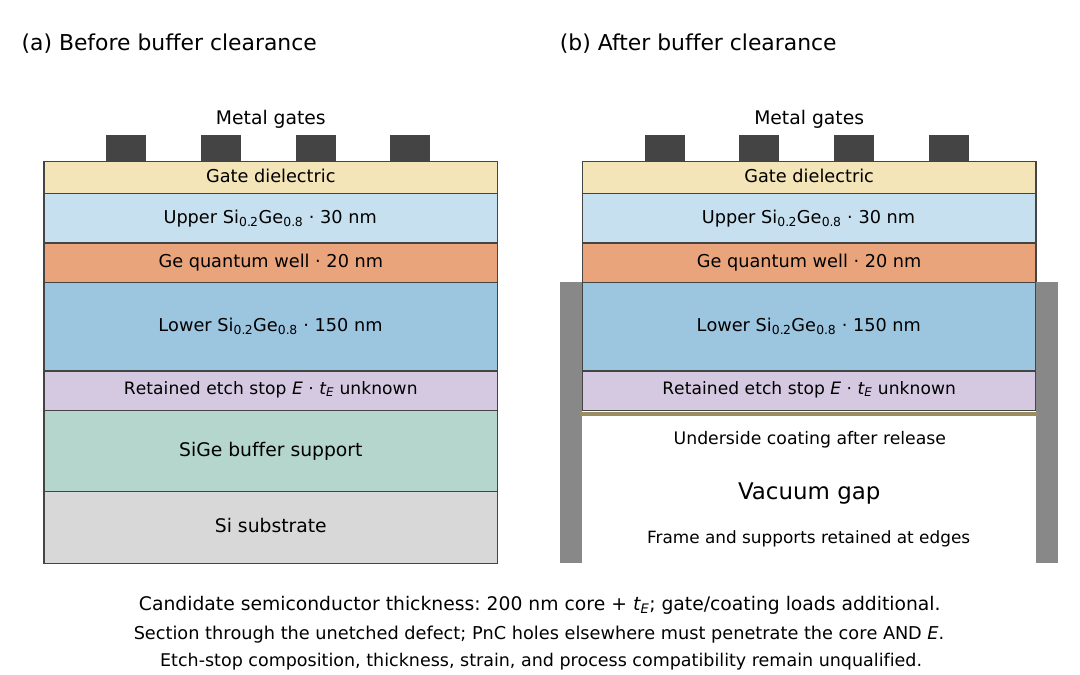}
\caption{Candidate layer retention before and after backside release,
not to scale. The 200 nm active core comprises 150 nm lower
Si$_{0.2}$Ge$_{0.8}$, 20 nm Ge, and 30 nm upper Si$_{0.2}$Ge$_{0.8}$.
The etch-stop layer $E$ is retained, so the semiconductor thickness is
$200~\mathrm{nm}+t_E$. Its composition and thickness, and the gate/coating
loads, remain inputs to be fixed and qualified. The released schematic
therefore specifies intended retention, not a finalized or fabricated
stack. PnC holes, omitted in this cross-section through the defect,
must penetrate both the core and $E$.}
\label{fig:heterostructure_stack}
\end{figure*}

The device concept in Secs.~\ref{sec:device_design}–\ref{subsec:operating_window} assumes a
high-mobility Ge 2DHG with strong and tunable spin--orbit coupling,
integrated with low-noise gate dielectrics and a suspended PnC membrane.
In this section, we specify a candidate SiGe/Ge stack whose suitability
must be checked after release, drawing on recent progress in Ge/SiGe epitaxy and Ge surface
passivation for quantum devices and advanced CMOS technology.%
~\cite{scappucci2021germanium,li2023quantumtransport,kong2024highmobility}
A schematic of the envisioned stack and its integration with the suspended PnC membrane is
shown in Fig.~\ref{fig:heterostructure_stack}.

\subsection{SiGe/Ge heterostructure}
\label{subsec:heterostructure}

The qubits reside in a compressively strained Ge quantum well grown on a relaxed SiGe virtual
substrate, which has emerged as a leading materials platform for planar Ge spin qubits and
high-mobility 2DHGs.%
~\cite{scappucci2021germanium,tanaka2012upper,myronov2007very,zudov2014miro}
A graded Si$_{1-y}$Ge$_y$ buffer on Si(001), followed by relaxed SiGe,
provides a virtual substrate for a strained Ge quantum well and SiGe
upper barrier in planar heterostructures.%
~\cite{adelsberger2019shallow,Ruggiero_NanoLett_2024_BackgatePlanarGe,Zhang_JOS_2024_GeSiGe}
The candidate retention route in Fig.~\ref{fig:heterostructure_stack}
additionally contains an epitaxial etch-stop layer $E$ between that
sacrificial buffer and the active core. From bottom to top, the retained
semiconductor layers are $E$, 150 nm Si$_{0.2}$Ge$_{0.8}$, 20 nm Ge,
and 30 nm Si$_{0.2}$Ge$_{0.8}$. Gates and gate dielectric lie above the
upper barrier; any underside or hole-wall coating is an additional
retained layer. The semiconductor thickness is therefore
$200~\mathrm{nm}+t_E$. The composition, thickness and growth compatibility
of $E$ remain design selections; the cited planar growth precedents do
not establish that complete suspended stack.

Compressive strain in the Ge well modifies the valence-band structure, enhancing the
heavy-hole character of the ground state, increasing the out-of-plane effective $g$-factor,
reducing hyperfine-induced dephasing, and enabling strong, electrically tunable spin--orbit
coupling favorable for EDSR-based control.%
~\cite{scappucci2021germanium,kloeffel2018theory}

State-of-the-art Ge/SiGe heterostructures grown by reduced-pressure chemical vapor deposition
(RPCVD) or molecular beam epitaxy (MBE) routinely achieve 2DHG mobilities above
$10^{5}$–$10^{6}~\mathrm{cm^2/Vs}$ at millikelvin temperatures and hole densities of order
$10^{11}~\mathrm{cm^{-2}}$,%
~\cite{tanaka2012upper,li2023quantumtransport,adelsberger2019shallow,kong2024highmobility}
These mobility measurements characterize transport disorder; they do
not by themselves establish an impurity concentration below
$10^{10}~\mathrm{cm^{-3}}$ in a nanometer-scale well. Chemical purity,
interface disorder, and charge noise require separate measurements on
the chosen wafer and processed device.

The structural quality of the virtual substrate and quantum well is typically assessed by
high-resolution x-ray diffraction, which confirms the strain state and layer thicknesses,
and by cross-sectional transmission electron microscopy, which reveals threading-dislocation
densities in the $10^{6}$–$10^{7}~\mathrm{cm^{-2}}$ range and abrupt, well-controlled
interfaces.%
~\cite{myronov2007very,adelsberger2019shallow}
These metrics are consistent with recent demonstrations of gate-defined Ge quantum dots and
planar Ge devices coupled to superconducting resonators,%
~\cite{hendrickx2020fast,lawrie2020quantum,li2023quantumtransport}
supporting the feasibility of the two-qubit, PnC-integrated architecture proposed here.
For first-generation devices, the same layout can be implemented in natural-abundance
Ge/SiGe to validate fabrication, tuning, and readout. For coherence-optimized devices,
however, the Ge quantum well and SiGe barriers should be grown from isotopically enriched,
nuclear-spin-depleted precursors, for example $^{70}\mathrm{Ge}/^{28}\mathrm{Si}^{70}\mathrm{Ge}$ or an
analogous spin-zero enriched-Ge platform such as one based on $^{74}\mathrm{Ge}$. This distinction
separates near-term process validation from the longer-term goal of testing
relaxation-limited two-qubit coherence in a high-purity, nuclear-spin-suppressed
environment.

\subsection{Surface passivation and gate dielectric}
\label{subsec:surface_chemistry}

The candidate gate stack uses an ALD dielectric on the upper SiGe barrier.
The exposed underside and PnC walls require separate qualification because
their compositions and process histories differ. Results on bare Ge,
Si$_{0.7}$Ge$_{0.3}$, and the proposed Si$_{0.2}$Ge$_{0.8}$ barriers are
not interchangeable measures of interface quality or spin dephasing.

Sulfur termination, plasma nitridation, and deliberate oxidation are
distinct preparation routes. Sioncke \emph{et al.} exposed cleaned Ge to
H$_2$S to produce sulfur termination. Subsequent TMA/H$_2$O ALD of
Al$_2$O$_3$ gave a Ge-oxide-free interface in that study, whereas the
TMA/O$_3$ route formed a GeO$_x$S$_y$ interlayer; H$_2$S treatment should
therefore not be described as forming GeO$_x$.~\cite{sioncke2011s}
Sardashti \emph{et al.} studied NH$_3$ plasma nitridation of
Si$_{0.7}$Ge$_{0.3}$, finding that a 300$^\circ$C pretreatment favored
Si-containing nitride/oxynitride passivation and suppressed Ge
out-diffusion.~\cite{Breeden_JAP_2020_LowTempThermalALD_DielectricOnSemiconductor}
The separate N$_2$/H$_2$ downstream-plasma study by Breeden \emph{et al.}
also concerns Si$_{0.7}$Ge$_{0.3}$.~\cite{breeden2019al2o3}
Neither process establishes a transferable recipe for the completed
Ge-rich suspended gate stack.

Slow traps in deliberately oxidized Al$_2$O$_3$/GeO$_x$/Ge stacks depend
on the process and layer structure.~\cite{ke2019slowtraps}
We therefore assign no numerical interface-trap density to the proposed
device from those precedents. Witness capacitors and transistors should
measure capacitance--voltage dispersion, leakage, hysteresis, and bias
stress before the same process is tested for charge noise in the dots.
Al$_2$O$_3$ and HfO$_2$ remain candidate dielectrics: the former has a
relative permittivity of order 9, while the latter provides a larger
permittivity; neither guarantees low noise simply by being high-$\kappa$.
~\cite{isometsa2019al2o3,delabie2005hfo2}
A nominal 5--15 nm gate-dielectric range is a design choice whose leakage,
capacitance, stress, and compatibility with release must be measured.

Passivation quality also depends on the observable. The a-Si:H/Al$_2$O$_3$
stacks studied by Berghuis \emph{et al.} reduced surface recombination in
Ge, but that result is not a measurement of spin coherence in a suspended
Ge/SiGe quantum dot.~\cite{berghuis2021passivation}
The gate and underside treatments will be selected using the actual
surface chemistry and electrical noise measurements, with coating loads
included in the mechanical model.

\section{Nanofabrication Process Flow}
\label{sec:fabrication}

\begin{figure}[t]
\centering
\includegraphics[width=\linewidth]{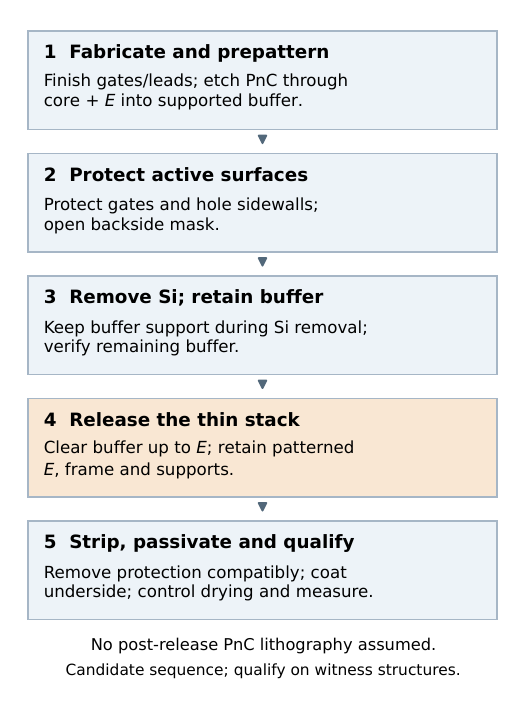}
\caption{Candidate process order. PnC patterning is completed while the
core and etch stop $E$ are supported by the buffer/substrate. Holes
penetrate $E$ as well as the core. After front/sidewall protection and
backside Si removal, clearing the remaining buffer up to $E$ is the
final thin-stack release step; $E$ remains in the membrane. Earlier Si
removal creates a buffer-backed composite that can also relax. No post-release
PnC lithography is assumed. Materials, protection/removal chemistry,
stresses, and gate survival require qualification on witness structures.}
\label{fig:process_flow}
\end{figure}

We outline a candidate sequence for integrating planar gates with a
released PnC, while identifying the unresolved etch-stop, protection,
and geometry choices. Its ordering is explicit; chemical compatibility,
yield, and final acoustic performance remain to be demonstrated.%
~\cite{hendrickx2020fast,lawrie2020quantum,li2023quantumtransport}
The sequence adapts planar gate-defined quantum-dot processing to the
candidate module in Sec.~\ref{sec:device_design}, with additional
patterning, protection, and release steps for the suspended PnC.%
~\cite{arrangoiz2019resolving,chu2017quantumacoustics,bienfait2019phonon}
A schematic overview of the key fabrication steps is shown in
Fig.~\ref{fig:process_flow}.

\subsection{Quantum dot and gate patterning}
\label{subsec:qd_gates}

Fabrication begins with definition of active mesas and ohmic contacts on the
SiGe/Ge heterostructure wafer. Optical lithography combined with dry or wet
etching is used to isolate $50$–$200~\mu$m-scale mesas that host individual
devices.%
~\cite{lawrie2020quantum,li2023quantumtransport}
Ohmic contacts to the 2DHG, serving as source, drain, and reservoir regions, are
formed by depositing a metal stack such as Pt/Al or Ti/Al, followed by a
forming-gas anneal. This step promotes diffusion into the Ge quantum well and
yields low-resistance p-type contacts compatible with cryogenic operation.%
~\cite{li2023quantumtransport,kong2024highmobility}

After mesa and contact formation, the exposed upper
Si$_{0.2}$Ge$_{0.8}$ barrier requires a surface-preparation process qualified
for its composition and the selected gate dielectric, as discussed in
Sec.~\ref{subsec:surface_chemistry}. Cleaning and oxidation conditions
will be selected using witness structures rather than transferred from
a bare-Ge recipe. A candidate $5$--$15~\mathrm{nm}$ ALD Al$_2$O$_3$ or
HfO$_2$ layer then provides the gate insulator. Its interface passivation,
leakage, and charge-noise performance require electrical qualification.%
~\cite{breeden2019al2o3,Breeden_JAP_2020_LowTempThermalALD_DielectricOnSemiconductor,scappucci2021germanium}

The quantum-dot gates are patterned using high-resolution electron-beam
lithography (EBL) with a positive-tone resist (e.g., PMMA or ZEP) or a bilayer
stack optimized for liftoff.%
~\cite{hendrickx2020fast,lawrie2020quantum}
The gate layout includes plunger gates controlling the electrochemical potential
of each dot, barrier gates tuning tunnel couplings to reservoirs and between the
two dots, and auxiliary screening or accumulation gates that shape the local
electrostatic environment. A typical metal stack is Ti/Pd/Au or Ti/Pt/Au,
deposited by electron-beam evaporation to ensure low surface roughness and good
step coverage. Ti provides adhesion to the dielectric, Pd or Pt acts as a stable
diffusion barrier, and the Au capping layer minimizes series resistance at low
temperature. After liftoff, the resulting fine gates have linewidths and spacings
in the $20$--$50~\mathrm{nm}$ range. The reference dot spacing is
$100~\mathrm{nm}$; lithographic dimensions do not directly determine
the wavefunction size or guarantee independent control
(Sec.~\ref{subsec:gate_qubits}).%
~\cite{hendrickx2020fast}

In addition to the plungers and barrier, the revised layout includes
separately wired local field or shape gates $S_1,S_2$ (Fig.~\ref{fig:device_concept}).
Their electrostatic effects will generally overlap. A self-consistent
three-dimensional Poisson--multiband calculation is needed before mask
finalization; the one-dimensional calculation here is a localization
test, not that electrostatic solution. Experimentally, all cross responses
enter the virtual-gate calibration in Sec.~\ref{subsec:charge_tuning};
the number of electrodes alone does not establish the required control rank.

\subsection{PnC patterning and membrane release}
\label{subsec:pncfab}

We specify a candidate process order, with the remaining materials choices
identified explicitly. The active semiconductor core comprises a 150 nm
Si$_{0.2}$Ge$_{0.8}$ lower barrier, a 20 nm Ge well, and a 30 nm
Si$_{0.2}$Ge$_{0.8}$ upper barrier. An additional epitaxial etch-stop layer,
denoted $E$, is retained beneath the lower barrier in this candidate route.
Its composition, thickness $t_E$, coherent strain, and selectivity have not
been established. The final semiconductor thickness is therefore
$t_{\mathrm{sem}}=200~\mathrm{nm}+t_E$, not 200 nm. Metal gates and
dielectric coatings add further patterned mechanical loads. Retaining $E$
resolves its intended fate; it does not establish a growth-compatible or
qualified final stack. The core-only calculations in
Sec.~\ref{subsec:membrane_stability} must be extended to include $E$ before
they can be used for this process.

\paragraph{Candidate sequence and the mechanical release step.}
The process order in Fig.~\ref{fig:process_flow} is as follows.
\begin{enumerate}
\item Form the mesas, contacts, gate dielectrics, gate electrodes, and
RF/DC wiring while the active layers remain supported by the SiGe buffer
and Si substrate. Define the backside mask before final release.
\item Pattern the PnC while that support is still present. Etch each hole
through the complete intended suspended semiconductor stack, including
$E$, and slightly into the underlying sacrificial buffer as needed for
etch completion. The dot-bearing defect and its intended connections to
the frame remain unetched. Stopping holes above $E$ would leave a continuous
layer across them and would produce a different acoustic structure.
\item Apply a qualified temporary protection scheme to the gate stack
and the exposed semiconductor hole sidewalls. Open the backside window
and remove the underlying Si, leaving sufficient SiGe buffer to support
the patterned active layer during this stage. Removing Si creates a
buffer-backed suspended composite, whose stress can already redistribute;
it does not yet create the intended thin membrane. A selective Si etch
alone does not define the final membrane thickness.
\item Remove the remaining buffer locally from below, stopping at $E$
and preserving the designed frame and supports. Clearing this supporting
buffer is the final thin-stack release step. The patterned core and the
patterned, retained $E$ layer then form the suspended semiconductor
structure. No subsequent PnC lithography on a released membrane is assumed.
\item Remove temporary protection by a process compatible with the gate
dielectric, contacts, and exposed semiconductor; passivate the now
accessible underside; and use a compatible stiction-controlled drying
sequence whenever a wet step follows release. The final coating geometry,
residual stress, charge, and mechanical loss enter the device model.
\end{enumerate}

This sequence is a proposal to qualify on witness structures, not a
demonstrated recipe. In particular, the sacrificial-buffer etchant must
preserve $E$, the protected hole sidewalls, and the completed gate stack;
the protection-removal step must not strip the gate dielectric. Chen
\emph{et al.} demonstrated pre-patterning and sidewall protection followed
by separate Si undercut and SiGe-buffer removal in Ge/SiGe multiple-quantum-
well microdisks, with an inserted etch-stop layer remaining in the
released structure~\cite{chen2017microdisks}. Their released $6$--$10~\mu\mathrm{m}$ microdisks were independently characterized
by Raman spectroscopy, photoluminescence, and 3D elastic modeling.
Their ungated optical device does not qualify the same wet chemistry or
mask removal for our gate stack.
Backside release integrated with micrometer-thick Si/SiGe spin-qubit
structures provides a further process precedent~\cite{marcogliese2026membranes},
but does not validate this much thinner Ge-well device.

The finite defect, shield, supports, and retained-layer dimensions must be
fixed before choosing the release window or predicting its mechanical
response. Lithographic placement must align both dots with the relevant
strain components, as defined in Eq.~\eqref{eq:strain_overlap}; a
displacement antinode alone is insufficient. Thickness and sidewall
metrology, gate-survival tests, and released strain/deflection measurements
are required to evaluate the candidate sequence.

\subsection{Core-only strain-sharing reference and its limits}
\label{subsec:membrane_stability}

Local compression in the Ge well, net membrane tension, and out-of-plane
curvature are distinct quantities. Elastic strain sharing on release has
been measured in Si/SiGe nanomembranes~\cite{roberts2006strainsharing}.
We calculate that effect for a coherent, unperforated, freely relaxing
Si$_{0.2}$Ge$_{0.8}$/Ge/Si$_{0.2}$Ge$_{0.8}$ core of total thickness
$t_0=200~\mathrm{nm}$ and Ge-well thickness 20 nm. This reference excludes
the retained layer $E$, gates, coatings, supports, holes, residual growth
stress, and differential thermal contraction. It is consequently neither
a model of the candidate released device nor a bound on its strain or bow.
The symmetric 90/20/90 nm reference is a comparison with the shallow-well
150/20/30 nm core, not an alternative fabricated structure.

Room-temperature lattice references are $a_{\rm Ge}=5.658$~\AA\ and
$a(x)=5.431+0.20x+0.027x^2$~\AA\ for Si$_{1-x}$Ge$_x$
\cite{dismukes1964lattice}. The assumed biaxial moduli are
$M_{\rm Ge}=140.5~\mathrm{GPa}$ and $M_{\rm SiGe}=148.5~\mathrm{GPa}$,
from $M=C_{11}+C_{12}-2C_{12}^2/C_{11}$ and linearly interpolated
room-temperature elastic constants~\cite{wortman1965elastic}.
With bulk Ge as the lattice reference, layer $i$ has stress-free
eigenstrain $e_i^*$ and elastic strain
$e_i(z)=e_0+\kappa z-e_i^*$. Free force and moment balance gives
\begin{equation}
\begin{pmatrix}A&B\\B&D\end{pmatrix}
\begin{pmatrix}e_0\\\kappa\end{pmatrix}
=\begin{pmatrix}
\sum_i M_i e_i^*(z_i-z_{i-1})\\
\frac12\sum_i M_i e_i^*(z_i^2-z_{i-1}^2)
\end{pmatrix},
\label{eq:laminate_balance}
\end{equation}
where $A=\sum_iM_i(z_i-z_{i-1})$,
$B=\tfrac12\sum_iM_i(z_i^2-z_{i-1}^2)$,
$D=\tfrac13\sum_iM_i(z_i^3-z_{i-1}^3)$, and $z=0$ is the core mid-plane.
For the SiGe barriers $e_i^*=-0.008788$, while
$e_{\mathrm{Ge}}^*=0$. The coherent as-grown Ge compression in this
reference is approximately $-0.879\%$.

\begin{table}[t]
\centering
\caption{Free-laminate results for the semiconductor core only. Thicknesses
are lower SiGe/Ge/upper SiGe. The last column is the chord displacement
$|\kappa|L^2/8$ at an illustrative $L=5~\mu\mathrm{m}$, not a calculated
deflection of an anchored PnC membrane. The retained etch stop $E$ and all
patterned-device loads and boundaries are excluded.}
\label{tab:laminate_reference}
\begin{tabular}{lccc}
\hline
Core (nm) & Ge strain & $|\kappa|$ ($\mathrm{m}^{-1}$) & Chord (nm)\\
\hline
$90/20/90$ & $-0.795\%$ & $0$ & $0$\\
$150/20/30$ & $-0.704\%$ & $1.51\times10^4$ & $47$\\
\hline
\end{tabular}
\end{table}

The symmetric core retains $90.5\%$ of its initial Ge compression and
has zero preferred curvature within this idealization. The asymmetric
core retains $80.1\%$ and has natural curvature
$1.51\times10^4~\mathrm{m}^{-1}$. The corresponding geometric chord
displacements are 47 nm at 5 $\mu$m and 189 nm at 10 $\mu$m. These are
reference chord lengths, not dimensions of a finalized release window.
The actual boundary-constrained shape, including possible buckling,
requires finite-device elasticity. The equal strain at two dots at the
same depth is imposed by lateral uniformity in this calculation; it
does not demonstrate equal local strains in the patterned device.
Nanoscale measurements of a supported Ge two-qubit device found
gate-associated strain variations of order $2\times10^{-4}$, illustrating
the importance of the omitted local gate geometry~\cite{corleywiciak2023strain}.

The core mid-plane expansion is about $8.4\times10^{-4}$ relative to the
as-grown SiGe lattice. A hypothetical uniform dilation of all acoustic
lengths at fixed material properties would give
$\delta f/f=-\delta a/a$, or about $-5$ MHz at 6 GHz. Release is not such
a uniform dilation: thickness, curvature, prestress, and local mode shape
change separately. This number is only a scaling illustration, not a
calculated contribution or bound on the cavity-frequency shift. We do
not infer a device detuning budget or recoverable frequency range from it.

To answer the device-specific strain and frequency questions, the next
model must include the selected $E$ layer, all gate/coating layers, and the
finite perforated membrane with its actual supports. A geometrically
nonlinear static calculation, checked for stable equilibrium, should
report the deflection and full strain tensor at each dot after release
and cooling. A prestressed acoustic eigenproblem on that relaxed geometry
then determines the mode frequency, strain overlap, and radiation leakage.
AFM or interferometry and Raman or X-ray nanodiffraction can constrain
the corresponding deflection and strain inputs. No actual two-dot strain
map, released-device bow, or cavity-frequency shift is established here.

\begin{figure}[t]
\centering
\includegraphics[width=0.98\linewidth]{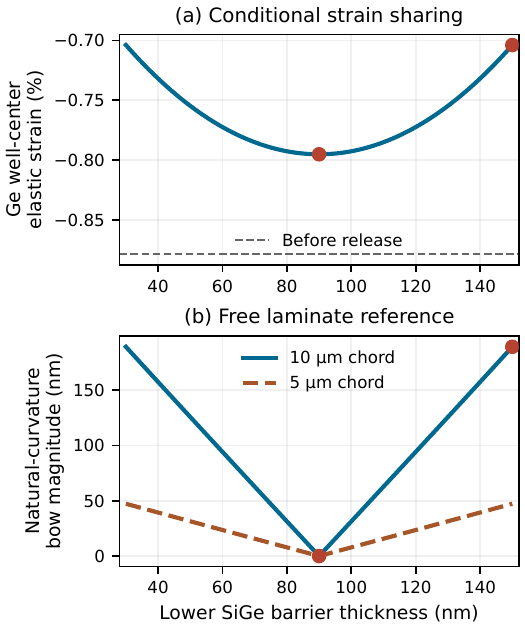}
\caption{Core-only free-laminate calculation at fixed 200 nm thickness and
20 nm Ge-well thickness. The upper barrier is 180 nm minus the lower
barrier thickness. (a) Ge well-center elastic strain; the dashed line is
the coherent as-grown value. (b) Geometric displacements for the natural
curvature and the indicated illustrative chord lengths. Highlighted cases
are in Table~\ref{tab:laminate_reference}. Neither panel includes the
retained etch stop $E$ or predicts the actual membrane shape or quality
factor.}
\label{fig:membrane_reference}
\end{figure}
\paragraph{Experimental Ge/SiGe membrane analogue and limits of transfer.}
Chen \emph{et al.} fabricated suspended Ge/SiGe multiple-quantum-well
microdisks using selective substrate removal and a retained etch-stop
layer~\cite{chen2017microdisks}. Their investigated microdisks had lateral
diameters of $6$ and $10~\mu\mathrm{m}$ and reported structural thicknesses
of $340$ and $440~\mathrm{nm}$. Three-dimensional elastic finite-element
modeling, Raman line scans, and photoluminescence measurements established
that strain redistribution depends on disk dimensions, layer configuration,
and externally applied SiN stressor geometry. In particular, adding an
all-around high-stress SiN film produced as much as $0.88\%$ \emph{additional
biaxial tensile-strain transfer} in their thinnest, smallest investigated
disk. This value is \emph{not} a measurement of the intrinsic Ge compression
remaining after release without a stressor and must not be inserted into
our compressively strained Ge-well model as an additive release strain.
The reported bending direction and spatial variation in the stressed disk
support the need for a nonuniform mechanical calculation; no transferable
numerical bow amplitude is provided for our anchored PnC device.

The relevance and limitation of this comparison are quantitative:
the published released structures establish a demonstrated Ge/SiGe
etch-stop fabrication route, strain changes measurable by Raman/PL,
and a realistic sensitivity to geometry and retained films. Our
$200~\mathrm{nm}$ SiGe/Ge/SiGe core is thinner, includes a different
well layout and gate/dielectric loading, and is patterned with PnC holes
rather than a plain optical disk. Accordingly, the measured $0.88\%$
externally induced strain is evidence of mechanical sensitivity, \emph{not}
a prediction of the strain at either of our dots. The optical resonances
reported for those microdisks are not GHz acoustic defect modes; their
measured shifts cannot be used as an acoustic-frequency correction. As
an explicitly illustrative scale check only, if a $6~\mathrm{GHz}$
acoustic eigenfrequency followed $f\propto L^{-1}$ under an assumed
uniform in-plane size change $\delta L/L=8.4\times10^{-4}$ with fixed
elastic properties, $\delta f\simeq-5~\mathrm{MHz}$; the actual
prestressed acoustic frequency requires recomputing the mode on the
relaxed, perforated, anchored geometry. Thus the literature analogue
supports the proposed release strategy and demonstrates potential
strain sensitivity, while the conditional core-only calculation above
supplies a separate transparent numerical reference. Neither is a
validated finite-device cavity-frequency prediction.

\subsection{RF and DC wiring}
\label{subsec:wiring}

Before membrane release, lithography and metallization define the RF and
DC wiring required
to control and read out the two-qubit module. On-chip coplanar waveguides (CPWs)
with a characteristic impedance of $50~\Omega$ are patterned to route microwave
signals to selected gates for electric-dipole spin resonance and to proximal
charge sensors or resonators used for qubit readout.%
~\cite{hendrickx2020fast,petit2020universal}
The CPWs typically employ a thick Au metallization with a Ti adhesion layer to
minimize attenuation and provide robust thermal anchoring at cryogenic
temperatures.

DC wiring for gate biases and source–drain contacts is integrated on chip or
routed through a custom printed-circuit board (PCB) to which the device is
wire-bonded. Large-area Au bond pads around the chip periphery facilitate
reliable Al or Au wire bonding to the chip carrier used in a dilution refrigerator
or a $1$–$4~\mathrm{K}$ cryostat. Ground planes and, where required, air bridges
are incorporated to suppress slot-line modes and minimize RF cross-talk,
following established best practices from spin-qubit and superconducting-qubit
architectures.%
~\cite{krinner2019engineering,petit2020universal}
Together, these wiring and packaging steps complete the nanofabrication flow and
prepare the device for low-temperature characterization of the phonon-mediated
two-qubit interactions introduced in Sec.~\ref{subsec:pncbus}.

\section{Readout Architecture and Signal Chain}
\label{sec:readout}

\begin{figure}[htp]
    \centering
    \includegraphics[width=0.90\linewidth]{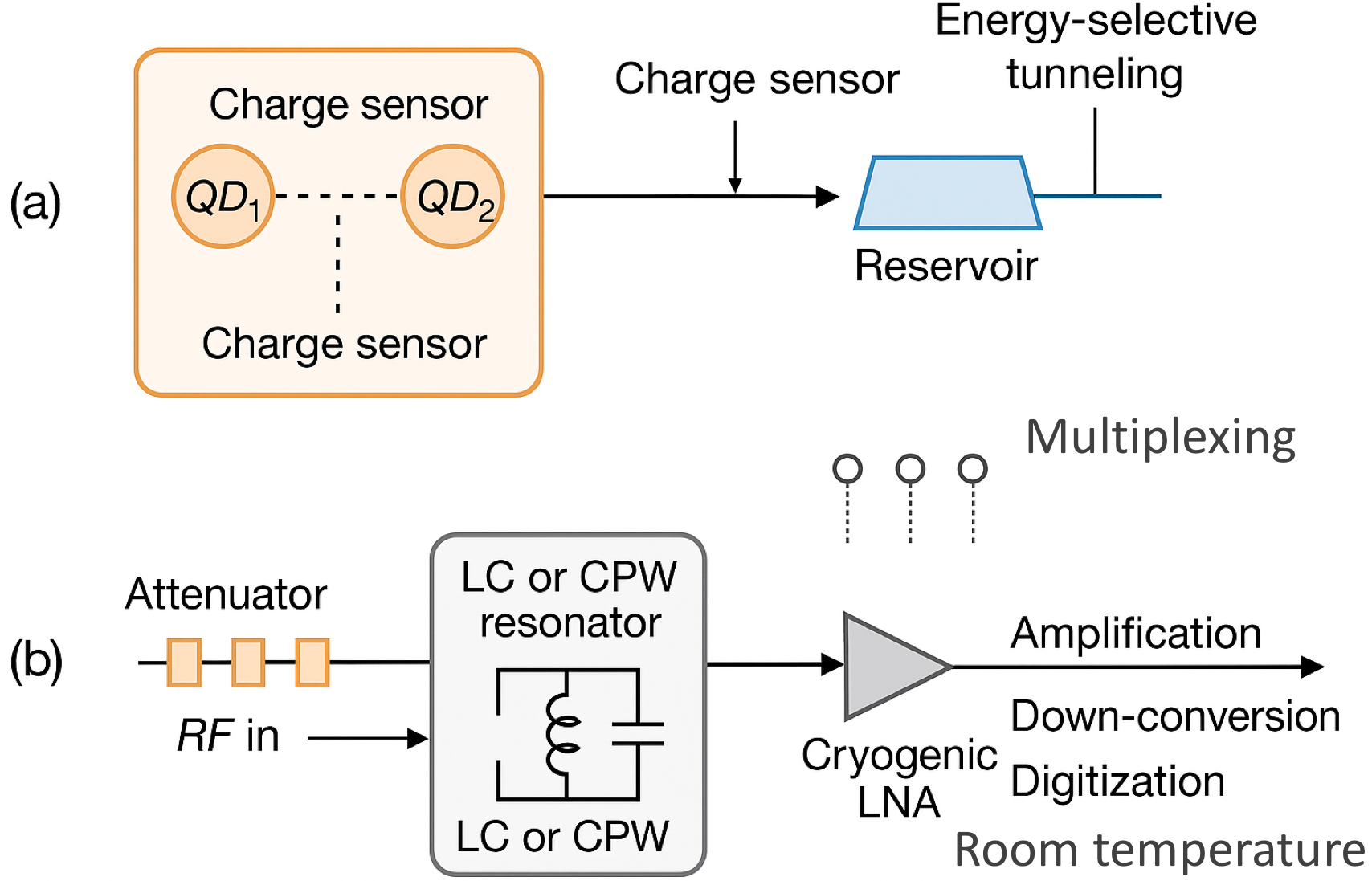}
    \caption{Illustrative readout architecture and signal chain.
    (a) Double-dot qubit pair (QD$_1$, QD$_2$) coupled capacitively to a nearby
    charge sensor (sensor QD or QPC). Spin-to-charge conversion is performed either
    by energy-selective tunneling to a reservoir or via Pauli spin blockade between
    the two dots.
    (b) The charge sensor is embedded in an RF resonant circuit (LC or CPW
    resonator) and probed by reflectometry. The input line includes distributed
    attenuators for thermalization, while the output line passes through a
    cryogenic low-noise amplifier (LNA) before further amplification, down-conversion,
    and digitization at room temperature. Multiple resonators can be frequency
    multiplexed on a single RF line in future multi-qubit devices.}
    \label{fig:readout_chain}
\end{figure}

The candidate flow in Sec.~\ref{sec:fabrication} is intended to provide a
two-qubit module with an integrated charge sensor, RF coplanar waveguides,
and DC control lines. This section describes the proposed readout architecture.
Our approach follows well-established spin-to-charge conversion protocols developed
for GaAs, Si, and Ge spin \mbox{qubits},%
~\cite{elzerman2004single,ono2002current,johnson2005triplet,hendrickx2020fast,petit2020universal}
and uses RF reflectometry of a proximal sensor dot or quantum point
contact (QPC) to target high-bandwidth charge sensing. Spin-to-charge
conversion and single-shot spin-readout fidelity must be qualified at
the intended $1$--$4~\mathrm{K}$ operating temperature.%
~\cite{field1993noninvasive,reilly2007fast,colless2013rfqd,gonzalezzalba2015singleshot}
An overview of the qubit, sensor, and RF chain is shown schematically in
Fig.~\ref{fig:readout_chain}.

\subsection{Spin-to-charge conversion}
\label{subsec:spin_to_charge}

The readout sequence converts the spin state of each hole-spin qubit into a charge
configuration detectable by the nearby sensor, closely following protocols demonstrated
for single-electron and single-hole spins in III--V and group-IV quantum dots.%
~\cite{elzerman2004single,ono2002current,johnson2005triplet,morello2010singleshot,hendrickx2020fast}
Two standard approaches are envisioned:

\begin{enumerate}
  \item \emph{Energy-selective tunneling (EST)}:%
  ~\cite{elzerman2004single,morello2010singleshot}
  during readout, a gate pulse rapidly shifts the electrochemical potential of the
  dot such that only one Zeeman branch (e.g., $\ket{\uparrow}$) is resonant with a
  nearby reservoir while the other branch ($\ket{\downarrow}$) remains below the
  Fermi level. Within a fixed readout window, an $\uparrow$ spin can tunnel out
  and be replaced by an opposite-spin hole (or the dot can empty, depending on
  bias), whereas a $\downarrow$ spin remains trapped. For a tunnel-out/reload protocol, the measured signal is the
  time-resolved charge transient: a reloaded dot and a dot that never emptied
  have the same final occupation. The sensor bandwidth and detection window
  must resolve the empty interval, and the tunnel-out, reload, and spin-relaxation
  rates must be calibrated. A final-occupation discriminator instead requires
  an explicitly latched mapping that retains different charge states. At $4$--$8$ GHz and $1$--$4$ K, the Zeeman splitting is smaller
  than $k_BT$, so this elementary energy filter does not by itself provide
  high-fidelity spin discrimination. Its thermal error must be measured;
  a lower temperature or an independently demonstrated amplified or
  measurement-based conversion protocol is required if it is inadequate.

  \item \emph{Pauli spin blockade (PSB)}:%
  ~\cite{ono2002current,johnson2005triplet,petta2005coherent,hendrickx2020fast}
  here the two-dot system is tuned near a charge transition such as
  $(1,1)\leftrightarrow(0,2)$, where only singlet states can form the $(0,2)$
  configuration. A gate pulse maps the joint spin state onto either a “blocked’’
  configuration (triplet, remaining in $(1,1)$) or an “unblocked’’ configuration
  (singlet, transitioning to $(0,2)$). The resulting charge configuration is then
  detected by the sensor as a change in electrostatic potential or current. PSB provides singlet--triplet discrimination, rather than complete
  readout of all four two-spin basis states in one measurement. Rotated PSB measurements can test Bell-state correlations, but
  local rotations of a singlet projector alone do not determine the
  individual Bloch vectors. Full state or process tomography requires
  informationally complete, calibrated readout mappings: individual
  spin-to-charge conversion, a qualified reference-spin or sequential
  scheme, or additional calibrated entangling pre-rotations.
  Its warm-operation window depends on the measured singlet--triplet
  gap, tunneling rates, and relaxation. PSB has been demonstrated in
  Ge hole double dots.%
  ~\cite{hendrickx2020fast}
\end{enumerate}

In both schemes, the spin-relaxation time $T_1$ in the relevant charge configuration
must exceed the measurement integration time to avoid relaxation-induced errors;%
~\cite{elzerman2004single,morello2010singleshot,yoneda2018quantum}
this requirement informs the design of tunnel couplings, Zeeman energies, and
phononic environment described in Secs.~\ref{subsec:pncbus} and~\ref{subsec:operating_window}.
The charge trajectory, or the latched charge configuration when such a
mapping is implemented, is sensed with a proximal detector sensitive to
the addition or removal of a single hole.

\subsection{Charge sensor and RF reflectometry}
\label{subsec:rf_reflectometry}

The charge sensor is implemented either as a quantum point contact (QPC) or as
an auxiliary quantum dot (sensor dot) located within $\sim 100~\mathrm{nm}$ of the
qubit pair. QPC-based charge sensing was first demonstrated in GaAs quantum dots,%
~\cite{field1993noninvasive} 
and has since been adapted to Si and Ge platforms,%
~\cite{li2023quantumtransport,gonzalezzalba2015singleshot}
while sensor dots offer enhanced sensitivity and tunability in strongly confined
heterostructures.~\cite{colless2013rfqd,west2019rfmultiplexed}

To reach the high bandwidth and signal-to-noise ratio required for single-shot
readout, the sensor is embedded in an RF resonant circuit and probed using
RF reflectometry.~\cite{reilly2007fast,colless2013rfqd,gonzalezzalba2015singleshot}
In the simplest implementation, the sensor resistance in series (or parallel)
with a lumped inductor forms an $LC$ resonator with a resonance frequency
$f_0 = 1/(2\pi\sqrt{LC})$ typically in the $100$--$500~\mathrm{MHz}$ range.
An RF tone near $f_0$ is injected through a directional coupler on the input line.
Changes in the sensor conductance shift the resonant frequency and quality factor,
which manifest as changes in the amplitude and phase of the reflected signal
$S_{11}(f)$.

The reflected tone is amplified and demodulated, and the resulting quadrature
signals are integrated over an optimized time window to discriminate the two
charge states. This RF-dot or RF-QPC approach routinely achieves sub-$\mu\mathrm{s}$
charge-detection times and single-shot spin readout in a variety of platforms.%
~\cite{reilly2007fast,colless2013rfqd,gonzalezzalba2015singleshot,west2019rfmultiplexed}
In the present architecture, the resonator and sensor are laid out in the
same metallization step as the CPWs and bond pads (Sec.~\ref{subsec:wiring}),
ensuring a compact footprint and low parasitic inductance.

Frequency multiplexing of several resonators on a single RF line is possible by
assigning distinct resonance frequencies (separated by a few to tens of MHz) to
different sensors and probing them simultaneously with a multi-tone input.%
~\cite{hornibrook2014multiplexed,west2019rfmultiplexed}
This capability will be important for scaling beyond the two-qubit module, but the
basic hardware is already compatible with the readout scheme described here.

\subsection{RF link budget and charge discrimination}
\label{subsec:link_budget}

We estimate the integration time $\tau_m$ for an idealized RF charge
discriminator under representative 1--4 K conditions. A 5\% charge
classification error and an additional SNR margin are considered below.
This link budget does not predict spin-readout fidelity: spin-to-charge
conversion, thermal activation, relaxation, event-detection bandwidth,
and state preparation must enter a separate measured error budget.

Unlike millikelvin operation, the primary noise contribution at $4~\mathrm{K}$
is the system noise temperature, dominated by the thermal noise of the device
and the noise-equivalent temperature of the cryogenic amplifier chain. For a
conservative baseline, we take a total system noise temperature
$T_{\text{sys}} \approx 8~\mathrm{K}$, comprising approximately
$4~\mathrm{K}$ device noise and approximately $4~\mathrm{K}$ added noise from
the LNA and cabling.

For dispersive RF readout, the signal-to-noise ratio (SNR) relates to the reflected signal power difference $\Delta P_{\text{sig}}$ (the effective signal power contrasting the two charge signals) and the integration time $\tau_m$ via:
\begin{equation}
    \text{SNR} = \frac{\Delta P_{\text{sig}} \cdot \tau_m}{k_B T_{\text{sys}}}
\end{equation}
For an input RF power of $-95$ dBm and reflection-amplitude contrast
$|\Delta\Gamma|=0.3$, $P_{\rm in}|\Delta\Gamma|^2=-105.46$ dBm
before output loss and detection inefficiency. We assume an additional
$9.54$ dB effective loss to obtain the illustrative amplifier-input
signal power $\Delta P_{\rm sig}=-115$ dBm
($3.16\times10^{-15}$ W). This expression is intended as an energy-resolution estimate; a device-specific readout model should also include resonator bandwidth, impedance matching, quadrature-detection efficiency, amplifier gain calibration, and the measured charge-state separation in the $I$--$Q$ plane. 

Using the explicitly chosen Gaussian discriminator convention
$\epsilon=\tfrac12\operatorname{erfc}(\sqrt{\mathrm{SNR}}/2)$,
$\epsilon=5\%$ corresponds to $\mathrm{SNR}=5.41$. We use
$\mathrm{SNR}=13$ as a conservative charge-discrimination margin
($\epsilon\simeq0.54\%$ in this idealized model), not as the threshold
for $95\%$ spin readout. Spin-to-charge conversion, relaxation, and
state-preparation errors remain separate.
Solving for the integration time:
\begin{equation}
\begin{aligned}
\tau_m 
&= \frac{\mathrm{SNR}\, k_B T_{\mathrm{sys}}}{\Delta P_{\mathrm{sig}}} \\
&\approx \frac{13 \cdot \left(1.38 \times 10^{-23}\,\mathrm{J/K} \cdot 8\,\mathrm{K}\right)}
{3.16 \times 10^{-15}\,\mathrm{W}} \\
&\approx 0.45 \times 10^{-6}\,\mathrm{s}.
\end{aligned}
\end{equation}

This yields a nominal integration time of $\tau_m \approx 450~\mathrm{ns}$.
With a more conservative signal level of
$\Delta P_{\text{sig}} \approx -125~\mathrm{dBm}$, the estimate increases to
$\tau_m\approx4.5~\mu\mathrm{s}$. These integration times must be
compared with measured relaxation in the readout configuration. For
example, $4.5~\mu\mathrm{s}$ is already a substantial fraction of the
$6.87~\mu\mathrm{s}$ illustrative cavity-induced relaxation time at
$4$ K in Sec.~\ref{subsec:operating_window}. The readout pulse may use a
different detuning and barrier, but its own spin-to-charge conversion,
thermal activation, and relaxation errors must be measured. The
amplifier-noise estimate alone is not a spin-readout fidelity. The link budget therefore defines an experimentally testable readout
target; it does not replace device-specific calibration using charge-state
histograms, noise spectra, and measured $T_1$ during readout.
\subsection{Cryogenic RF chain}
\label{subsec:cryo_chain}

The RF reflectometry signal chain follows best practices developed for semiconductor
spin qubits and superconducting qubits.%
~\cite{reilly2007fast,krinner2019engineering,petit2020universal}
On the input side, the RF drive is heavily attenuated and filtered at multiple
temperature stages (e.g., 300~K, 40~K, 4~K, and base temperature) to both set the
desired power at the device and thermalize noise from room-temperature electronics.
Typical configurations use $40$--$60~\mathrm{dB}$ of total attenuation, split across
stages to avoid excessive local heating.

On the output side, the reflected RF signal passes through an isolator or circulator
to protect the device from back-propagating amplifier noise, then is amplified by a
cryogenic low-noise amplifier (LNA) mounted at the 4~K stage or, in higher-performance
setups, at the base plate. Commercial LNAs operating in the $50$--$500~\mathrm{MHz}$
range achieve noise temperatures of a few kelvin with gains of $30$--$40~\mathrm{dB}$,%
~\cite{weinreb2007ultralownoise,krinner2019engineering}
which is compatible with our $1$--$4~\mathrm{K}$ operating window and modest drive powers.

After further amplification and filtering at room temperature, the signal is
down-converted in an IQ mixer referenced to the same local oscillator used to
generate the probe tone. The resulting in-phase ($I$) and quadrature ($Q$)
components are digitized by a fast ADC and processed in real time (e.g., in an FPGA)
or off-line on a computer to extract the spin state via thresholding or maximum-likelihood
classification.%
~\cite{colless2013rfqd,west2019rfmultiplexed}
The same RF chain can be straightforwardly extended to multiple sensors through
frequency multiplexing, as mentioned above.

\subsection{Readout acceptance measurements}
\label{subsec:readout_performance}

The target of greater than $95\%$ single-shot spin readout is conditional.
First, calibrated charge-state histograms and the resonator response must
establish the effective signal power, noise, and bandwidth. Second,
state-preparation tests must separately determine spin-to-charge
conversion errors, thermal leakage, relaxation, and excitation during
the measurement pulse. A sensitive charge sensor does not resolve a
thermally broadened spin conversion window by itself.

Within a linear sensor response and fixed detection convention, the
integration time decreases as the squared charge-state signal separation
increases and increases with the system noise. Resonator bandwidth and
transient response must still be included. Incident RF power is not
equivalent to locally dissipated power, so a nominal input below
$-90$ dBm does not establish negligible heating. Drive-power and duty-cycle
sweeps, charge thermometry, and qubit/mode spectroscopy will determine
an acceptable readout point. Spin fidelity and local heating remain
measured acceptance criteria, not consequences of the link budget.
\section{Cryogenic Measurement Setup}
\label{sec:measurements}

\begin{figure}[t]
    \centering
    \includegraphics[width=0.90\linewidth]{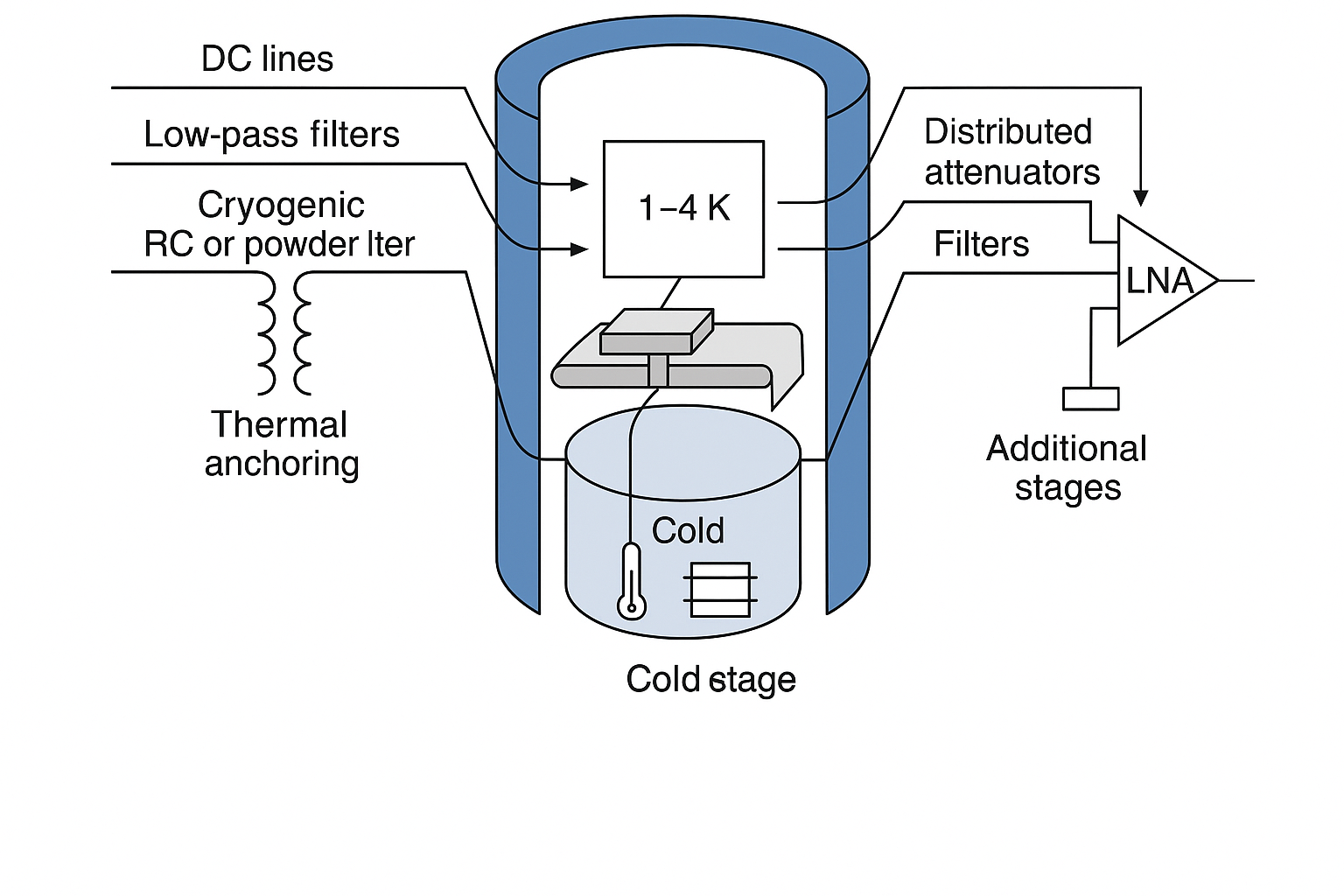}
    \caption{Illustrative cryogenic measurement setup for the Ge-based two-qubit device.
    The chip is wire-bonded to a PCB or ceramic carrier mounted on the cold stage of a
    $1$--$4~\mathrm{K}$ cryostat inside a superconducting magnet.
    DC gate and bias lines (left) pass through room-temperature low-pass filters,
    cryogenic RC or powder filters, and thermal anchoring stages before reaching the chip.
    RF lines for EDSR and reflectometry (right) are routed through distributed
    attenuators and filters on the input side and through a cryogenic low-noise
    amplifier (LNA) and additional stages on the output side.
    Thermometers and heaters on the sample stage provide temperature control and
    stabilization over the $1$--$4~\mathrm{K}$ range.}
    \label{fig:measurement_setup}
\end{figure}

The readout architecture described in Sec.~\ref{sec:readout} is implemented in a
cryogenic measurement environment that provides low-noise electrical access,
a well-controlled magnetic field, and stable operation in the $1$--$4~\mathrm{K}$
temperature window targeted for this Ge hole-spin platform. The setup follows
best practices established for semiconductor spin qubits and superconducting
circuits,%
~\cite{reilly2007fast,krinner2019engineering,hornibrook2014multiplexed}
with wiring, filtering, and thermal anchoring tailored to RF reflectometry and
phonon-mediated control. A schematic overview is shown in
Fig.~\ref{fig:measurement_setup}.

The fabricated chip (Sec.~\ref{sec:fabrication}) is mounted on a gold-plated copper
PCB or a low-loss ceramic carrier and wire-bonded to the DC and RF bond pads.
The carrier is then thermally anchored to the cold stage of a cryostat capable of
reaching $1$--$4~\mathrm{K}$, such as a pumped $^4$He system or a closed-cycle $^4$He
cryostat with a low-vibration sample plate.%
~\cite{petit2020universal,yang2020operation,krinner2019engineering}
To minimize microphonic pickup and ensure robust thermalization, the carrier is
clamped or bolted with high-conductivity interfaces (e.g., indium gaskets or
thermally conductive foils), and bond-wire lengths are kept as short as practical.

DC gate voltages and source/drain biases are supplied by low-noise voltage sources
at room temperature. Their outputs pass through feedthrough $\pi$-filters or RC
low-pass filters at the cryostat entry, followed by additional cryogenic filtering
(e.g., lumped-element RC filters, copper-powder filters, or lossy coaxial lines)
anchored at intermediate stages (40~K and 4~K) before reaching the sample.%
~\cite{lukashenko2008cryofilters,bluhm2011overhauser,reilly2007fast}
This staged filtering suppresses broadband electromagnetic interference and reduces
low-frequency charge noise that would otherwise broaden the qubit resonance and
limit $T_2^*$ and readout fidelity.

RF signals for EDSR and reflectometry readout
(Sec.~\ref{sec:readout}) are routed using semi-rigid coaxial cables with
well-characterized attenuation and thermal conductivity.%
~\cite{reilly2007fast,krinner2019engineering}
On the input side, cascaded attenuators (typically totaling $40$--$60~\mathrm{dB}$)
are distributed across temperature stages (300~K, 40~K, 4~K, and the cold stage) to
thermalize room-temperature noise and set the desired RF power at the device while
limiting dissipation on the cold plate. To suppress out-of-band noise and spurious
mixer products, absorptive (e.g., Eccosorb) or commercial band-pass filters can be
placed close to the sample.%
~\cite{krinner2019engineering}

On the output side, the reflected RF signal from the sensor resonator
(Sec.~\ref{subsec:rf_reflectometry}) passes through an isolator or circulator
anchored at 4~K to reduce back-propagating amplifier noise, and is then amplified
by a cryogenic LNA with a noise temperature of a few kelvin.%
~\cite{weinreb2007ultralownoise,bryerton2013ultralownoise}
Subsequent amplification, filtering, and IQ down-conversion are performed at room
temperature as described in Sec.~\ref{subsec:cryo_chain}, after which the signal is
digitized and processed to infer the charge (and hence spin) state.

A static magnetic field is provided by a superconducting solenoid or vector magnet
surrounding the sample space. The field magnitude and orientation are selected to
exploit the anisotropic and gate-tunable $g$-factor in Ge hole-spin qubits while
limiting unwanted orbital effects that can degrade confinement and stability.%
~\cite{hendrickx2020fast,scappucci2021germanium,kloeffel2018theory}
The field is selected from the measured $g$-tensor of each dot.
The earlier fixed field interval does not guarantee the required
frequencies for arbitrary in-plane orientations; joint resonance is
established by the calibration in Sec.~\ref{subsec:charge_tuning}.%
~\cite{petit2020universal,yang2020operation}

Finally, calibrated thermometry (e.g., Cernox or RuO$_2$ sensors) and resistive
heaters mounted near the sample enable closed-loop temperature stabilization and
controlled sweeps across the $1$--$4~\mathrm{K}$ range.%
~\cite{krinner2019engineering}
These capabilities are essential for mapping temperature-dependent coherence and
readout performance and for validating that the engineered phononic environment
remains stable under realistic drive and operating conditions. Together, the
cryogenic infrastructure, wiring, and magnetic-field control provide a complete
experimental platform for characterizing the phonon-mediated two-qubit operations
proposed in this work.

\section{Experimental Program and Target Metrics}
\label{sec:benchmarks}

\begin{figure*}[t]
\centering
\includegraphics[width=0.97\textwidth]{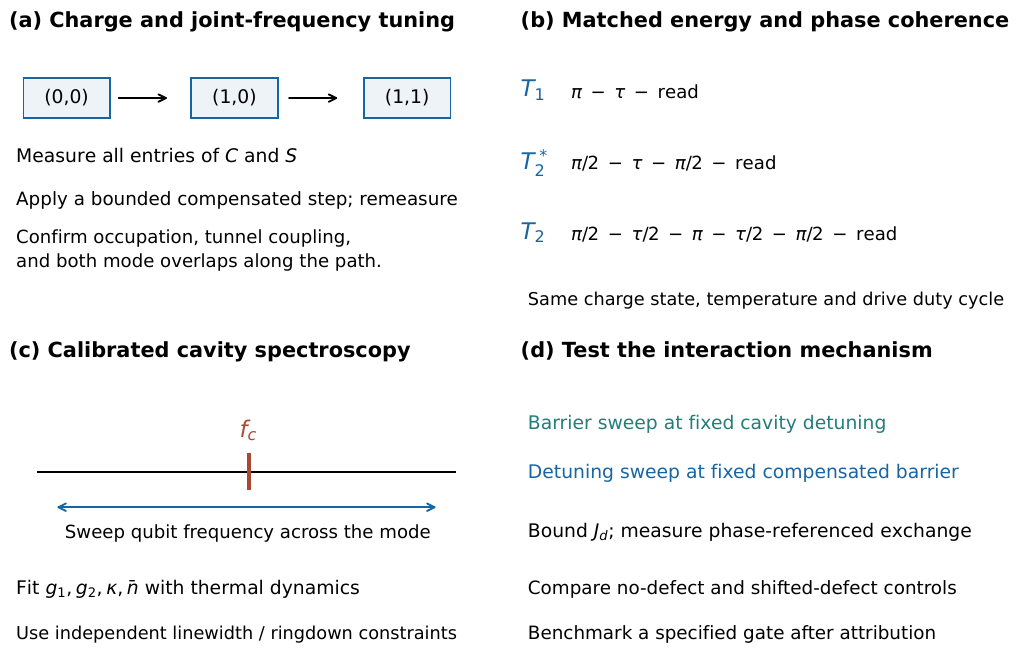}
\caption{Proposed measurement sequence; no experimental data are shown.
(a) Establish single-hole occupation and calibrate cross-compensated
frequency controls. (b) Measure population relaxation, Ramsey dephasing,
and Hahn echo at matched operating points. (c) Fit both qubits and the
cavity with thermal occupation and independent linewidth constraints.
(d) Separate direct and cavity-mediated exchange using barrier sweeps,
detuning sweeps, phase references, and control devices before evaluating
the fidelity of a specified gate.}
\label{fig:experimental_program}
\end{figure*}

We take the spin--phonon parameters of Ref.~\cite{mei2025qst} as
conditional modeling inputs and translate them here into concrete design rules
and experimentally testable benchmarks for a two-qubit module. The cryogenic
infrastructure in Sec.~\ref{sec:measurements} and the readout architecture in
Sec.~\ref{sec:readout} provide the key ingredients needed to execute this program.
Our measurement sequence proceeds from charge tuning and noise baselining, through
single-qubit control and PnC validation, and culminates in spectroscopic
and time-domain signatures of two-qubit coupling. The workflow follows established
protocols developed in GaAs, Si, and Ge spin-qubit experiments,%
~\cite{elzerman2004single,petta2005coherent,morello2010singleshot,hendrickx2020fast,petit2020universal}
but is structured to isolate and quantify the role of the engineered phononic
environment. Figure~\ref{fig:experimental_program} summarizes the proposed measurements,
while the corresponding quantitative targets---guided by Ref.~\cite{mei2025qst}---are
summarized in Table~\ref{tab:targets}.

\subsection{Charge stability and tuning}
\label{subsec:charge_tuning}

The first stage is to establish robust electrostatic control of the double dot
in the few-hole regime and to verify reliable RF readout via the proximal charge
sensor. We will adapt standard charge-sensing methods developed for GaAs and Si
double dots%
~\cite{field1993noninvasive,elzerman2004single,colless2013rfqd}
to the Ge/SiGe platform by sweeping pairs of gate voltages while monitoring the
demodulated RF reflectometry signal
(Sec.~\ref{subsec:rf_reflectometry}). The resulting charge stability diagrams should
exhibit the characteristic honeycomb pattern of a double dot, with well-resolved
addition lines corresponding to changes in the occupations of QD$_1$ and QD$_2$. The charge-loading sequence is summarized in Fig.~\ref{fig:experimental_program}(a).

An early milestone is to reach the $(0,0)$ charge state and then step into the
$(1,0)$, $(0,1)$, and $(1,1)$ configurations, confirming single-hole control in each
dot, as demonstrated in prior Si and Ge spin-qubit experiments.%
~\cite{hendrickx2020fast,lawrie2020quantum,morello2010singleshot}
From these data we will extract gate lever arms, charging energies, and inter-dot
capacitive coupling, which in turn define operating points for single-qubit control
and for PSB readout
(Sec.~\ref{subsec:spin_to_charge}). Charge-noise levels can be quantified by tracking
temporal fluctuations of selected charge transitions, yielding an effective charge-offset
noise amplitude and spectral density that can be compared across devices and materials
systems.%
~\cite{bluhm2011overhauser,freeman2016comparing}
These electrostatic and noise metrics provide the baseline against which the coherence
and phonon-engineering benchmarks in Table~\ref{tab:targets} will be evaluated.

\paragraph{Joint tuning with charge and barrier compensation.}
A common magnetic field supplies a coarse frequency adjustment;
unequal local $g$ tensors prevent it from setting both qubits
independently. The candidate layout therefore includes separately wired
local field or shape gates $S_1,S_2$, in addition to $P_1,P_2$ and
$B_{12}$. Their drawn presence is not evidence of independent control.
Virtual-gate compensation is established experimentally, and tunnel
cross-talk can be calibrated through its logarithmic gate
dependence.~\cite{hendrickx2024sweetspot,hsiao2020orthogonal}
The required rank, useful voltage range, and retained coupling for this
released module have not yet been calculated or measured.

At fixed magnetic-field orientation and magnitude, write
$\mathbf V=(V_{P_1},V_{P_2},V_{B_{12}},V_{S_1},V_{S_2})^T$,
$\mathbf q=(\mu_1,\mu_2,\ln(t_c/t_{c,\mathrm{ref}}))^T$, and
$\mathbf f=(f_1,f_2)^T$. The tunnel matrix element $t_c$ has units of
energy; $f_i$ are qubit frequencies obtained from a fit that accounts
for cavity-induced shifts. The local response is
\begin{equation}
 \delta\mathbf q=C\,\delta\mathbf V,\qquad
 \delta\mathbf f=S\,\delta\mathbf V.
 \label{eq:virtual_gate_jacobian}
\end{equation}
All off-diagonal derivatives are required. $C$ is determined from
charge-transition slopes, lever arms, and tunnel spectroscopy, while
$S$ requires measurements of both qubits under each independent gate
perturbation. Neither matrix can be inferred from the scalar orbital
potential in Sec.~\ref{subsec:gate_qubits}.

Conditioning must be assessed after specifying voltage budgets and
observable tolerances. Let the positive diagonal matrices $D_V,D_q,D_f$
contain, respectively, allowed gate-step scales, charge/barrier
tolerances, and frequency scales. Define
$\widehat C=D_q^{-1}CD_V$ and
$\widehat S=D_f^{-1}SD_V$, and choose orthonormal columns of $N$ to span
$\ker\widehat C$. A dimensionally consistent compensated local step is
\begin{equation}
 \delta\mathbf V
 =D_VN(\widehat S N)^+D_f^{-1}\delta\mathbf f,
 \label{eq:joint_virtual_step}
\end{equation}
where $+$ is the pseudoinverse. Independent frequency control requires
$\operatorname{rank}(\widehat S N)=2$ with singular values large enough
relative to calibration uncertainty and the available voltage range.
If the proposed step exceeds a voltage or linearization limit it must
be reduced, or a constrained solve must be used; the pseudoinverse
alone does not enforce those bounds. For three independent constraints
in $\mathbf q$, five controls leave two local directions. Once both
frequencies are set, no additional direction is generally available to
optimize $g_1$ or $g_2$. Their acceptance inequalities may already hold,
but independent repair of a failed overlap is not guaranteed.

\paragraph{Finite trajectory and charge margin.}
A local Jacobian is insufficient to establish a $150$ MHz tuning range.
For each measured cavity offset, the required evidence is a continuous
gate path $\mathbf V(s)$ from the initial $(1,1)$ state to the target
frequencies, with absolute biases, voltage limits, and residual errors
reported. At each step one must update $C,S$, the charge boundaries,
$t_c$, and the cavity frequency, reduce the step until the linearized
prediction agrees with the measurement or device calculation, and
verify that the full pulse excursion remains inside the charge cell.
For exchange, the dressed qubit frequencies must coincide within the
interaction linewidth, including unequal Lamb and mean thermal Stark
shifts. Correcting those means does not remove fluctuating differential
Stark shifts when $g_1\ne g_2$; the thermal dynamical model is still
required.

Keeping two local charge potentials fixed does not by itself preserve
the $(1,1)$ state: charging energies, mutual capacitance, and reservoir
alignment may also change. Let $\Delta E_a(s)>0$ be the measured or
calculated grand-energy cost of a relevant competing charge sector
relative to $(1,1)$, including charge-transfer sectors $(2,0)$ and
$(0,2)$ and addition/removal sectors. In the weak-charge-hybridization
limit, a conservative equilibrium charge-error check for a specified
budget $\epsilon_{\rm ch}$ is
\begin{equation}
 \sum_a r_a\exp[-\Delta E_a(s)/(k_BT_e)]
 \leq\epsilon_{\rm ch},
 \label{eq:charge_margin_test}
\end{equation}
where $T_e$ is the measured carrier temperature and $r_a$ includes the
relevant state multiplicities relative to $(1,1)$. The costs must be
evaluated over the actual pulse excursion and with the measured
broadening. Virtual charge admixture, of order $(t_c/\Delta E_a)^2$,
must also be small; equilibrium margins do not certify nonadiabatic
loading or pulse errors. Charge sensing throughout the trajectory is
required.
The assumed $U_{\rm eff}$ used in the orbital example is not this
measured charge-addition spectrum.

\paragraph{Quantitative screening scales (not a tuning demonstration).}
A $150$ MHz frequency correction at a fixed illustrative $B=0.20$ T
would require an effective $g$-factor adjustment
$|\delta g|=h|\delta f|/(\mu_B B)=0.05359$ for each qubit
requiring that shift, before dressing and cross-talk are included.
This is a necessary Zeeman scale, not a physical gate-voltage estimate:
the actual local derivatives $\partial g_i/\partial V_j$, their
cross terms, the allowed gate excursions and the cavity-induced shifts
remain unknown. Likewise, for just one competing charge sector with
unit relative multiplicity, the equilibrium weight condition in
Eq.~\eqref{eq:charge_margin_test} at $\epsilon_{\rm ch}=0.01$ requires
$\Delta E\ge k_BT_e\ln100$, i.e., $0.397$ meV at $1$ K and
$1.587$ meV at $4$ K. More competitors, state degeneracies,
charge hybridization and pulsed operation can tighten these conditions.
The pair of screening numbers cannot replace the device-specific
$(1,1)$ stability diagram or a compensated finite gate-voltage path.

\paragraph{Constructive five-strip electrostatic trajectory in a
specified ideal half-space (not a calibrated device).}
We evaluated a second exactly declared reduced physical-gate geometry:
two $35$ nm plunger strips at $x=\pm50$ nm, a $35$ nm central barrier
strip at $x=0$, and two $50$ nm screening strips at $x=\pm110$ nm.
The dot positions are $x=\pm50$ nm, sampled at depth $z=40$ nm.
For a uniform charge-free dielectric half-space with all unspecified
surface held at zero voltage, strip $j$ gives
\begin{align}
P_j(x,z)&=\frac1\pi\bigg[
\tan^{-1}\frac{x-x_j+w_j/2}{z} \nonumber\\
&\hspace{7mm}-\tan^{-1}\frac{x-x_j-w_j/2}{z}\bigg],\\
\delta\phi(x,z)&=\sum_j P_j(x,z)\delta V_j.
\label{eq:five_strip_poisson}
\end{align}
We use $E_z=-\partial_z\phi$, illustrative
$\partial g/\partial E_z=0.7\,(\mathrm{MV/m})^{-1}$,
$B=0.20$ T and $\partial[g_{\rm sp}/(2\pi)]/\partial E_z
=5.8$ MHz/(MV/m), with each unperturbed $g_{\rm sp}/2\pi=5$ MHz.
Solving a full five-by-five cross-response problem to hold
$\delta\phi(-50)=\delta\phi(50)=\delta\phi(0)=0$ while
imposing $\delta f_{1,2}=\pm150$ MHz yields the path
$\mathbf V(s)=s(10.519,-10.519,0,-21.895,21.895)$ mV
for $0\le s\le1$. Both couplings remain within
$[4.556,5.444]$ MHz in this \emph{assumed field-linear model}.
\begin{figure}[t]
\includegraphics[width=\columnwidth]{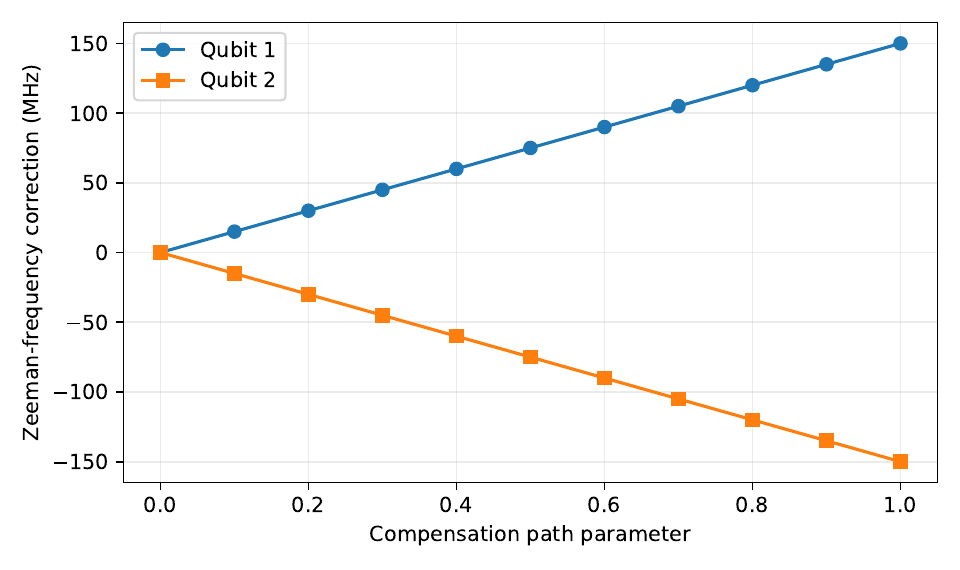}
\caption{Analytical half-space five-strip compensation path, showing
both calculated Zeeman corrections. The gate positions, linear
$g$-response and uniform dielectric are assumed rather than a full
self-consistent Ge/SiGe device computation.}
\label{fig:five_gate_path}
\end{figure}
For a deliberately chosen constant-interaction illustration
$U_1=U_2=8$ meV, $U_{12}=1$ meV and
$\mu_1=\mu_2=-4.5$ meV, the $(1,1)$ state has a minimum
competing-sector grand-energy gap of $3.5$ meV over the
\emph{ideal potential-preserving path}. Its equilibrium Boltzmann
sum over eight competing sectors at 4 K is $8.21\times10^{-5}$,
under the assumption of unit sector multiplicities and no tunnel
hybridization. This is a constructive \emph{existence example within
assumed parameters}: it does not calculate the charge sector of the
fabricated structure. In particular, vertical field-dependent orbitals,
barrier curvature, $t_c$, dielectric interfaces, screening, and
mode-shape-dependent couplings have not been solved self-consistently.
Thus it cannot be used to claim that the proposed device has a
qualified $\pm150$ MHz tuning range or remains in $(1,1)$.

\paragraph{Coupling acceptance and remaining validation.}
The same trajectory must provide $g_1(\mathbf V)$ and $g_2(\mathbf V)$,
from a multiband calculation with the released strain mode or from
resolved qubit--mode spectra fitted with thermal damping. Acceptable
points require both mode overlaps, a controlled direct-coupling bound,
the dispersive validity condition when that approximation is used, and
an exchange signal resolved relative to decoherence. The interaction
depends on $g_1g_2$, so a frequency solution near either strain node is
rejected. Spatially varying etch errors can change the mode shape and
cannot in general be repaired by a common frequency shift.

No device-calibrated cross-capacitance matrix, spin-frequency lever arm, or
$\pm150$ MHz charge-preserving trajectory is claimed here. The
schematic does not yet supply the full electrode geometry and material
boundary conditions needed to calculate them. Demonstrating a finite
trajectory with occupation, tunnel coupling, and both spin--phonon
couplings retained remains necessary to establish the proposed
module's retuning range and fabrication tolerance.

\subsection{Suspension and interface contributions to dephasing}
\label{subsec:dephasing_surface}

In this design, suspension is intended to reduce acoustic leakage into
the substrate and permit a PnC shield; it does not itself establish a
complete bandgap or a high-quality localized mode. Release exposes an
underside closer to the quantum well than the original substrate interface.
For the 150/20/30 nm core, the well center is 160 nm above the core base;
the retained etch-stop layer and coatings change the distance to the
actual exposed surface. Fluctuating interface charges can couple to the
spin through its electric-field-dependent properties. The resulting noise
must be measured for the chosen stack.~\cite{hendrickx2024sweetspot}

We propose evaluating a conformal underside dielectric after backside
access is available. The material exposed in the retained-layer route is
$E$, not necessarily Ge, and its surface treatment must be selected
accordingly. The chemical distinctions in Sec.~\ref{subsec:surface_chemistry}
also apply to any exposed SiGe and Ge walls. ALD can alter fixed charge,
trap noise, strain, and mechanical damping; a higher dielectric constant
or the use of TMA/H$_2$O alone does not establish beneficial Ge--O bonding
or improved spin coherence. Coating thickness, stress, hole-radius changes,
and loss must be included in the complete device model.

Matched supported/suspended devices, with and without the selected
underside treatment, will be compared using charge-noise spectra, Ramsey
and echo decays. These controls constrain changes associated with release
and passivation. A causal assignment also requires matched charge states,
drive conditions and temperatures, and calibration of altered strain,
confinement and acoustic modes. The comparison alone does not isolate a
single dephasing mechanism. In the backside-release route the buried
underside cannot be directly pre-passivated; the process therefore uses
post-access coating and a compatible stiction-controlled drying sequence.

\paragraph{Thermalization of the suspended membrane.}
For a small steady-state temperature rise, a local thermal model gives
$\Delta T\simeq P_{\rm diss}/G_{\rm th}$. Neither the dissipated power
nor the thermal conductance has been established for this device.
For illustration only, assuming $G_{\rm th}=10^{-8}$--$10^{-6}$ W/K
gives $\Delta T=1$--$100~\mu$K for $P_{\rm diss}=1$ pW and
$1$ mK--$0.1$ K for $1$ nW. Nonlinear thermal transport requires a
temperature-dependent conductance rather than this local estimate.

Heat can flow through modes outside the designed phononic gap, but
unpatterned tethers or metal straps can also bypass the acoustic shield
and increase cavity leakage. Their geometry must be included in both
the thermal and elastic loss models. Drive-dependent qubit spectra,
charge thermometry, and mode excitation/relaxation will constrain heating.
The mode temperature entering $\bar n$ is distinct from the amplifier
system-noise temperature entering the RF link budget.
\subsection{Single-qubit control and coherence}
\label{subsec:single_qubit_metrics}

With stable single-hole occupancy established, the planned single-qubit
characterization will use electric dipole spin resonance (EDSR), with
spin-to-charge readout as described in Sec.~\ref{subsec:spin_to_charge}.
At a fixed magnetic field, microwave-frequency drives will be applied to
a selected plunger or barrier gate, and the spin state will be read out
using a qualified energy-selective tunneling or PSB protocol.%
~\cite{elzerman2004single,ono2002current,hendrickx2020fast}
Resonance-frequency sweeps will determine the projected effective
$g$-factor of each dot at that field direction. Repeating this spectroscopy
versus magnetic-field orientation at each selected gate bias will
constrain the directional Zeeman splittings and the symmetric tensor
$\mathbf G_i=\mathbf g_i^{\mathsf T}\mathbf g_i$, through
$f_i^2=(\mu_B/h)^2\mathbf B^{\mathsf T}\mathbf G_i\mathbf B$.
A gate sweep at one fixed field direction measures only the tuning of
that projected splitting. Splitting measurements alone do not fix the
spin-basis rotation in $\mathbf g_i$ or the EDSR and spin--phonon matrix
elements; those require additional drive calibration or a validated
spinor model.%
~\cite{scappucci2021germanium,kloeffel2018theory}

Time-domain control experiments---Rabi oscillations, Ramsey fringes, Hahn echo, and, where useful,
Carr--Purcell--Meiboom--Gill (CPMG) sequences---will quantify coherence in the $1$--$4~\mathrm{K}$ window.%
~\cite{petta2005coherent,yoneda2018quantum,petit2020universal,hendrickx2020fast}
From these planned data we will extract the inhomogeneous dephasing time $T_2^\ast$, the echo coherence time
$T_2$, and the EDSR Rabi frequency for realistic drive amplitudes, as illustrated in
Fig.~\ref{fig:experimental_program}(b). Systematic sweeps versus temperature and field orientation
will help identify dominant decoherence mechanisms (e.g., charge noise, phonon-induced dephasing,
or residual hyperfine coupling) and quantify the benefit of the phonon-engineered environment.%
~\cite{maier2013relaxation,khaetskii2001spinrelax}

A practical near-term target for first-generation suspended devices is
$T_2^\ast \gtrsim 1~\mu\mathrm{s}$ and $T_2 \gtrsim 10$--$50~\mu\mathrm{s}$ at operating
points where single-qubit gate times (set by the Rabi frequency) lie in the
$10$--$100~\mathrm{ns}$ range. These are acceptance aims at separately qualified operating points.
For example, the single-spin weak-dispersive $4$ K example with
$T_1=6.87~\mu$s
has the Markovian relaxation limit $T_2\leq13.74~\mu$s; reaching
$50~\mu$s requires a different detuning, coupling, occupation, or loss
budget. Suspension and added interfaces can further reduce coherence.
The longer-term materials-motivated target is substantially more ambitious. For a single spin,
\begin{equation}
\frac{1}{T_2}=\frac{1}{2T_1}+\frac{1}{T_\phi},
\label{eq:t2_t1_tphi}
\end{equation}
for exponential Markovian decay at the same operating point, with
$T_1^{-1}=\Gamma_\downarrow+\Gamma_\uparrow$ at finite temperature.
The relaxation-limited condition $T_2=2T_1$ then corresponds to
$T_\phi^{-1}\ll (2T_1)^{-1}$. The reported $T_2=2T_1=1.2~\mathrm{ms}$ benchmark in
isotopically enriched Ge donor material therefore provides a useful upper-end materials
reference for high-purity, spin-zero-isotope Ge. Because that value was reported as a
relaxation-limited benchmark rather than as a full decomposition of all dephasing channels,
it does not provide a precise quantitative upper bound on $T_\phi^{-1}$, and the same
limit must still be tested in gate-defined Ge hole-spin devices.~\cite{sigillito2015electron}

For a two-qubit module, the relevant decay also depends on the
superposition being measured. In an exponential description, write
\begin{equation}
\Gamma_{2,\mathrm{2q}}\simeq\Gamma_{2,1}+\Gamma_{2,2}
+\Gamma_{\rm corr},\qquad \Gamma_{2,i}=T_{2,i}^{-1}.
\label{eq:twoqubit_t2}
\end{equation}
The single-qubit rates refer to the same coupled operating point.
$\Gamma_{\rm corr}$ describes covariance and interaction-dependent
effects not already included in those rates; it can have either sign
and vanishes for independent local noise. This avoids counting each
qubit's cavity-induced dephasing twice. For two identical, independently
dephasing spins, a Bell-like coherence has
$T_{2,\mathrm{2q}}=T_{2,\mathrm{1q}}/2$. Substituting the donor-material
reference $1.2$ ms gives $0.6$ ms only as a conditional arithmetic
example, not a bound or prediction for the hole-spin device. Neither
this ratio nor a count of nominal pulse durations establishes a gate
fidelity or usable circuit depth.
\subsection{PnC characterization}
\label{subsec:pnccavity_benchmarks}

Direct spectroscopy of GHz phononic modes is nontrivial in the present architecture, so we
characterize the PnC cavity primarily through its imprint on spin relaxation and
dephasing. The central measurement is the spin-relaxation time $T_1$ as a function of Zeeman
energy, obtained by sweeping the magnetic field while tuning the qubit transition frequency within,
across, and outside the designed PnC bandgap.%
~\cite{chu2017quantumacoustics,bienfait2019phonon,arrangoiz2019resolving,mei2025qst}
The spectroscopy protocol is summarized in Fig.~\ref{fig:experimental_program}(c).

A bandgap-related enhancement must be tested against matched unpatterned
and detuned-defect devices, not just against a different qubit frequency:
spin--orbit matrix elements and gate-noise sensitivity can also vary
with field and gate bias. In the weak, dispersive limit the defect adds
the frequency-dependent relaxation of Eq.~\eqref{eq:thermal_purcell}.
Near resonance, a strongly coupled, thermally occupied mode can generate
hybridized spectra and nonexponential dynamics; fitting a single $T_1$
dip then does not uniquely determine $g_i$ and $Q_m$.

Independent mechanical linewidth or ringdown information from matched
resonator witnesses, when available, is combined with both qubits' spectra
and upward/downward transition rates in a common thermal model. A
linewidth-based $Q_m=\omega_c/\kappa$ must use the energy-decay convention
of Sec.~\ref{subsec:operating_window}. Bandgap suppression and useful
two-qubit exchange are separate benchmarks: any enhancement of $T_1$
must be reported alongside $T_2^*$, $T_2$, mode occupation, and the
measured $g_1g_2$ at the interaction point.
\subsection{Two-qubit coupling signatures}
\label{subsec:twoqubit_signatures}

\paragraph{Direct and cavity-mediated coupling.}
We first increase the interdot barrier in compensated coordinates and
measure $t_c$ from the charge anticrossing or photon-assisted tunnelling.
For a symmetric, spin-conserving two-site Hubbard estimate with effective
charging energy $U_{\mathrm{eff}}$, the exchange energy is
$J_H\simeq4t_c^2/U_{\mathrm{eff}}$; its flip--flop angular coefficient is
$J_d\simeq J_H/(2\hbar)$. In a hole device, spin--orbit-induced anisotropy
and direct electrostatic effects must also be bounded experimentally.
Direct-interaction bounds require the quantitative residual-bus
control below and separate conditional spectroscopy of any longitudinal
term. A prespecified attribution criterion is that this upper limit
be below $10\%$ of the cavity contribution under test. This is a chosen
measurement margin, not a predicted device property. The barrier may be
pulsed to a different value for spin-to-charge readout after the interaction
interval.

\paragraph{Quantitative residual-bus control.}
Far-detuned exchange measurements constrain $J_{\rm tot}=J_d+J_c$,
not $J_d$ alone. A calibrated residual cavity contribution and its
uncertainty must be retained. If only a magnitude bound is available,
the conservative direct bound is
\begin{equation}
 |J_d|\leq |J_{\rm tot}|+|J_{c,\rm far}|,
 \label{eq:residual_direct_bound}
\end{equation}
with experimental and model uncertainties added to the right-hand side.
For $g_1/2\pi=g_2/2\pi=5$ MHz, detunings of $100$, $500$, and
$1000$ MHz give residual $|J_c|/2\pi$ of $250$, $50$, and $25$ kHz,
respectively. Thus $1$ GHz detuning still leaves a cavity term equal
to the example $25$ kHz direct-coupling margin. We will use a phase-referenced
joint fit at positive and negative detunings with independently
constrained $g_i$ and cavity parameters, or an independently qualified
coupling-off control. This prevents destructive cancellation from being
mistaken for suppression of the direct interaction. Both transverse
and longitudinal direct terms are bounded before claiming a full
cavity-mediated gate.

Next, both qubit frequencies will be swept together while preserving their
mutual resonance, $(1,1)$ occupation and calibrated tunnel coupling. For
each detuning we remeasure $g_i$, $t_c$, charge noise and local temperature.
In the weak-dispersive limit, the reactive response is
\begin{equation}
 J_c\simeq\frac{g_1g_2}{2}
 \sum_{i=1}^{2}\frac{\Delta_i}{\Delta_i^2+(\kappa/2)^2}.
 \label{eq:damped_cavity_exchange}
\end{equation}
For equal detuning it approaches $g_1g_2/\Delta$ and reverses sign across
the cavity when the coupling phases are held in a common reference. The
associated thermal relaxation follows the even-in-detuning response in
Eq.~\eqref{eq:thermal_purcell}. Spectroscopic splittings measure
$2|J_d+J_c|$ and do not determine the sign; sign-sensitive inference
requires phase-referenced exchange tomography. A barrier sweep at fixed
detuning tests the direct component, whereas detuning sweeps at fixed
barrier test the cavity component. Fits must allow the measured
gate-dependent $g_i$ and any residual direct term to vary within their
calibration errors.

An avoided crossing alone is insufficient evidence of phonon mediation.
We require agreement between the dispersive frequency dependence and
independently calibrated cavity parameters, a resolved time-domain exchange
signal, and controls on devices with no defect mode in the relevant band
or with a deliberately shifted defect. Near-resonance spectra are fitted
to Eq.~\eqref{eq:two_spin_bus} with thermal damping, rather than extending
Eq.~\eqref{eq:damped_cavity_exchange} through its invalid region.
The coupling rate, phase, damping and readout errors are then used for
process or interleaved randomized benchmarking of a specified gate;
frequency splittings or coherence-time ratios alone are not gate fidelities.

\paragraph{Relaxation and phase coherence at matched operating points.}
At each accepted frequency and temperature, we will acquire interleaved
$T_1$, Ramsey $T_2^*$ and Hahn-echo $T_2$ data on each qubit, with identical
charge state, field orientation, drive duty cycle and readout procedure.
Ground- and excited-state preparation will constrain upward and downward
transition rates separately; when a single-spin exponential description
is adequate, $T_1^{-1}=\Gamma_\downarrow+\Gamma_\uparrow$ describes return
to the measured equilibrium population. Matched unpatterned and
detuned-defect controls, together with calibration of gate-noise
sensitivity, mode occupation and heating, will test whether an observed
change is consistent with bandgap protection. The measured decay envelopes
and their fit exponents will be reported; no experimental $T_1$, $T_2^*$
or $T_2$ data for the proposed module are supplied in this study.
Only when an exponential Markovian description is adequate will we infer
$\Gamma_\phi=T_2^{-1}-(2T_1)^{-1}$; stretched or quasistatic Ramsey/echo
decays require an appropriate noise-spectrum analysis. Enhancement of
$T_1$ will therefore be tested independently of any enhancement of phase
coherence.

\paragraph{Collective decay and interpretation of individual rates.}
Equation~\eqref{eq:thermal_purcell} is the diagonal single-spin
weak-dispersive rate. When the two spins are mutually resonant, the
same cavity also produces correlated decay and excitation. For $\kappa\ll|\Delta_i|$, at leading
order in $g_i/\Delta_i$, these include
\begin{equation}
 L_- = \sqrt{\kappa(\bar n+1)}
       \sum_i\frac{g_i}{\Delta_i}\sigma_i^- ,\qquad
 L_+ = \sqrt{\kappa\bar n}
       \sum_i\frac{g_i}{\Delta_i}\sigma_i^+ .
 \label{eq:collective_purcell_jumps}
\end{equation}
The corresponding dissipators contain cross terms. Bright and dark
spin combinations and coherent exchange can therefore give
nonexponential local populations even far from the cavity resonance.
With the partner effectively decoupled, we will measure individual total
excitation and relaxation rates and record the changed working conditions.
If the cavity remains active, those rates still include its contribution.
Background rates will be inferred only after subtracting or jointly
fitting the calibrated cavity contribution, with the associated
uncertainties propagated. Alternatively, we will fit the full two-spin--mode
dynamics jointly. At the active exchange point we will report the prepared
state, population trajectories and state-specific coherences, rather than
identifying a local single-exponential fit with an intrinsic $T_1$.
The scalar $T_2^{-1}=(2T_1)^{-1}+T_\phi^{-1}$ relation will be used only
when its single-spin exponential assumptions are verified. The complete
coupled-mode equation includes the correlations in
Eq.~\eqref{eq:collective_purcell_jumps} without adding them a second time.
When the cavity is retained explicitly, additional spin Lindblad terms
represent background mechanisms only; total rates measured with the
cavity active cannot also be inserted as independent spin losses.
\begin{table*}[t]
\caption{Conditional targets and acceptance tests. None is demonstrated
for the proposed released Ge module. Numerical coupling and thermal values
are reduced-model examples, not a simultaneous device specification.}
\label{tab:targets}
\begin{ruledtabular}
\begin{tabular}{p{0.23\textwidth}p{0.28\textwidth}p{0.40\textwidth}}
Quantity & Reference or target & Required validation \\
\hline
Dot separation and oscillator length & $d=100$ nm, $\ell_x=20$ nm &
Self-consistent electrostatics, orbital spectroscopy, and measured $(1,1)$ charge cell \\
Membrane stack & 200 nm core plus retained etch stop $E$ &
Specify $E$, gates and coatings; validate release, local strain, bow,
and gate survival \\
Cavity frequency and quality factor & $f_c\sim5$--$6$ GHz;
$Q_m\gtrsim1.5\times10^4$ as a target &
Prestressed finite-device elastic model with converged radiation loss;
measured linewidth or ringdown \\
Spin--phonon couplings & MHz scale; illustrative $g_i/2\pi=5$--$10$ MHz &
Both $g_i$ calibrated along the joint-tuning trajectory; thermal model and no strain-node operating point \\
Thermal occupation at 6 GHz & $\bar n=3.00$ at 1 K; $13.40$ at 4 K &
Measure the effective mode temperature and excitation as well as relaxation \\
Relaxation & Bandgap protection tested at a specified detuning &
Example cavity-limited $T_1=27.3/6.87~\mu$s at 1/4 K for $g_i/2\pi=5$ MHz,
$\Delta_i/2\pi=100$ MHz, $Q_m=1.8\times10^4$ \\
Ramsey and echo coherence & Initial aims $T_2^*\gtrsim1~\mu$s,
$T_2\sim10$--$50~\mu$s &
Matched control devices; no inference of $T_2$ enhancement from $T_1$ alone \\
Cavity exchange and ideal duration & $J_c/2\pi=0.25$--$1$ MHz;
$t_{\sqrt{\rm iSWAP}}=0.125$--$0.5~\mu$s &
Residual direct coefficient $|J_d|<0.1|J_c|$;
quantified residual-bus, detuning and phase controls;
full thermal model and measured gate fidelity \\
Thermal phase channel & Specified plain and midpoint-echo sequences &
Appendix~\ref{app:thermal_channel} gives conditional channel diagnostics;
no device gate fidelity is predicted \\
Longer-term coherence projection & Donor-material reference $T_2=1.2$ ms;
independent-noise two-spin example $0.6$ ms &
Not a bound or prediction for hole spins; test correlations and cavity-induced phase noise \\
\end{tabular}
\end{ruledtabular}
\end{table*}
\section{Discussion and Outlook}
\label{sec:discussion}

The planar fabrication elements have precedents, but the integrated
released Ge/SiGe structure remains unqualified. Realizing the required phonon
$Q$-factors and reproducible defect-mode frequencies will still demand careful
process control. In particular, membrane thickness, feature dimensions (hole radii
and lattice constant), and etch-induced surface roughness all directly impact the
bandgap, scattering losses, and mode localization. Likewise, operation at
$1$--$4~\mathrm{K}$ places practical constraints on continuous RF drive: EDSR and
reflectometry powers must be chosen to achieve adequate signal-to-noise ratio
without introducing excessive local heating or destabilizing charge configurations.
These considerations are not unique to Ge, but they will set the operating envelope
for first-generation demonstrations based on the present module.

The experimental program in Sec.~\ref{sec:benchmarks} provides a stepwise path from
basic charge characterization to signatures of phonon-mediated two-qubit coupling.
The reduced-order results identify concrete decisions before fabrication.
The orbital calculations quantify how both separation and confinement
change direct tunneling; the new sensitivity analysis identifies
conditional parameter combinations that fail the chosen attribution margin. The laminate
calculation shows why Ge strain can remain compressive after release,
while an asymmetric stack can develop substantial natural curvature.
The taper calculation converts an etch angle into actual top and bottom
radii; it does not determine acoustic leakage. Consequently, full
device electrostatics, prestressed elastic eigenmodes, and a converged
$Q_m(\theta)$ calculation remain required. The specified controls turn
these open questions into tests, without treating a process plan as
demonstrated manufacturability.

\subsection{Interpretation and calibration of reduced-order feasibility studies}
\label{subsec:reduced_scope}
The added numerical studies answer distinct \emph{screening} questions; they
are not a single calibrated prediction for the released Ge/SiGe module.
For the five-strip electrostatic example in Sec.~\ref{subsec:charge_tuning},
a uniform charge-free half-space at 40 nm depth, prescribed field-to-$g$
and field-to-coupling slopes, and independently assigned constant-interaction
energies yield a $\pm150$ MHz compensation path using plunger increments
$\pm10.519$ mV and screening increments $\mp21.895$ mV. The modeled dot
and central barrier \emph{potentials} remain fixed, and the stipulated
couplings lie between $4.556$ and $5.444$ MHz. This is an existence
example for the stated response functions, not evidence that the actual
Ge hole orbitals, tunnel coupling, charge addition energies, or phononic
strain overlaps remain fixed along physical gate voltages. Those properties
require a self-consistent electrostatic, charge-state and multiband-hole
calculation using the selected gates and released structure.

The sidewall-angle sweep in Sec.~\ref{subsec:pncbus} uses a 21-site,
edge-damped \emph{one-dimensional} defect-chain surrogate rather than a
three-dimensional anisotropic elastic cavity. Its $Q_{\rm 1D}$ varies
from 44,824 at $0^\circ$ to 45,348 at $10^\circ$, with nonmonotonic behavior.
Varying the chain from 17 to 25 sites changes the zero-angle result
from approximately $9.33\times10^3$ to $2.16\times10^5$; this changes
the physical shielding length and is a \emph{geometry sensitivity}, not
mesh convergence. The surrogate's frequency is normalized to 6 GHz at
zero taper; its illustrative 10-degree change is approximately 542 MHz.
It must not be combined with the independently assumed $\pm150$ MHz
qubit-compensation range as a validated fabrication budget. Neither its
absolute $Q$ nor its frequency shift is a prediction for the proposed
three-dimensional Ge/SiGe device.

A physical-device qualification must first fix the retained etch-stop
material and thickness, full PnC defect and shielding geometry, tethers,
metal and dielectric loading, gate footprints, and relevant elastic and
band parameters. The same released geometry must then supply (i) a
prestressed open-boundary three-dimensional mode sweep with identified
mode and convergence of radiation-limited $Q(\theta)$ and (ii) calibrated
potential maps, orbital and charge-state energies, spin frequencies and
both spin--phonon couplings along \emph{one} finite compensated voltage
trajectory. These calculations remain outstanding. The present screening
studies constrain the next design iteration but do not demonstrate that
all referee-specified acceptance conditions hold in the fabricated module.

From a materials and device perspective, the two-qubit module leverages the key
advantages of strained Ge hole systems---large and electrically tunable spin--orbit
coupling, a proposed $1$--$4~\mathrm{K}$ characterization window, and Si-compatible
processing---while explicitly addressing historically challenging issues such as
charge noise and phonon-induced relaxation through optimized dielectrics and PnC
engineering.%
~\cite{scappucci2021germanium,berghuis2021passivation,arrangoiz2019resolving,mei2025qst}
A future demonstration of enhanced $T_1$ and robust $T_2$ within the PnC bandgap
would test the design choices in Secs.~\ref{subsec:pncbus}
and~\ref{subsec:pnccavity_benchmarks}, and could establish a transferable
strategy for phonon engineering in other spin-qubit platforms.

The proposed module would test whether the benefits of isotopic and
chemical purification survive in coupled gates and entangling operations.
At a single-spin operating point where exponential Markovian decay is
verified, $T_2\approx2T_1$ indicates a small additional pure-dephasing rate.
It does not establish phase protection for a two-spin state. Detuning,
tunnel coupling, exchange, cavity-frequency fluctuations and interfaces
introduced by suspension can affect the joint dynamics differently.
Matched single-spin relaxation and coherence measurements will test
Eq.~\eqref{eq:t2_t1_tphi}; joint population trajectories and state-specific
two-spin coherences will test whether the independent-noise description
in Eq.~\eqref{eq:twoqubit_t2} is adequate. Nonexponential or correlated
dynamics require the corresponding joint noise or channel model rather
than a single inferred pure-dephasing time. The donor reference is not an
upper bound for hole-spin coherence, and dividing a projected coherence
time by a single-qubit pulse duration does not establish usable two-qubit
circuit depth.

Several natural extensions follow from the present design. On the device side,
integrating superconducting microwave resonators or high-impedance transmission
lines with the PnC membrane would enable hybrid spin--phonon--photon architectures,
in which localized mechanical modes couple both to spins and to microwave photons
for long-range interconnects and multiplexed readout.%
~\cite{chu2017quantumacoustics,bienfait2019phonon,mi2018cavity,landig2019coherent}
Such hybrid devices could exploit the strong spin--phonon interaction of hole spins
together with electromechanical coupling in nanostructured Ge to access high-cooperativity
tripartite interactions, opening routes to phononic transduction and mechanically
protected quantum memories.

On the phonon-design side, alternative PnC motifs and defect geometries,
including snowflake lattices, bound states in the continuum and
topological edge modes, motivate studies of radiation leakage,
localization and fabrication sensitivity.%
~\cite{eichenfield2009optomechanical,tsaturyan2017ultracoherent,arrangoiz2019resolving}
A geometrical suppression of radiation loss would improve $Q_{\rm rad}$;
it does not by itself establish lower intrinsic material damping.
Changes in strain-energy participation or dissipation dilution require
their own stress and loss model. Comparisons of total $Q_m$ must therefore
identify radiation, intrinsic and loading contributions consistently.

Systematic comparison of measured $T_1(B)$ spectra with refined finite-element models
will guide these iterations, with the goal of simultaneously maximizing $T_1$ deep in
the bandgap while preserving MHz-scale spin--phonon coupling needed for fast entangling
operations.

At the system level, scaling beyond a single two-qubit module will require coordinated
progress in wiring, calibration, and architectural design. Linear chains or two-dimensional
arrays of phonon-coupled Ge qubits could be constructed by repeating the module and coupling
multiple cavities or defect modes on a shared membrane, while frequency-division multiplexing
of RF sensors and integration of cryogenic control electronics can mitigate wiring overhead.%
~\cite{hornibrook2014multiplexed,krinner2019engineering,vandersypen2017interfacing}
Achieving stable operation in larger arrays will also require cross-talk mitigation,
robust calibration workflows, and pulse sequences tailored to the decoherence budgets
identified in Secs.~\ref{subsec:single_qubit_metrics} and~\ref{subsec:twoqubit_signatures}.
These are possible extensions beyond the present two-qubit tests.

The proposed module is a candidate for joint mechanical, electronic, and
experimental validation. The present calculations establish conditional
constraints and reproducible diagnostics; they do not establish
manufacturability or a simultaneous warm operating point. Resolving the
finite geometry, retained-layer properties, gate-control range and thermal
channel is necessary before assigning a device performance specification.

\section{Conclusion}
\label{sec:conclusions}

We have presented a conditional design study for phonon-coupled Ge hole-spin
qubits that proposes integrating a strained Ge/SiGe heterostructure, a
suspended PnC cavity, gate-defined double quantum dots, and
RF-reflectometry-based readout. Building on recent advances in Ge materials and
spin-qubit platforms that demonstrate fast, high-fidelity control in planar
Ge/SiGe heterostructures,%
~\cite{scappucci2021germanium,hendrickx2020fast,li2023quantumtransport}
our design specifies a candidate materials stack, process order, and
measurement program. The new reduced-order calculations and calibrated
control requirements make the design assumptions explicit. The final
mask layout, released strain distribution, loss budget, and coherent
performance still require device-specific validation.
A central element of the architecture is the integration of the qubits into a
defect cavity of a two-dimensional PnC patterned directly into the Ge membrane.
Inspired by advances in optomechanical and quantum-acoustic systems,%
~\cite{eichenfield2009optomechanical,chu2017quantumacoustics,arrangoiz2019resolving,bienfait2019phonon}
the design objectives are to suppress unwanted phonon emission using a
PnC stopband and retain a localized mode with useful spin--phonon
coupling. These are properties to establish for a specified full stack
and finite geometry; neither a complete bandgap nor the requisite local
couplings follows from the present schematic. We
have outlined a comprehensive experimental program that employs established
spin-to-charge conversion and RF sensing techniques%
~\cite{elzerman2004single,colless2013rfqd,petit2020universal}
to characterize charge stability, single-qubit coherence, phonon-modified
relaxation, and ultimately phonon-mediated two-qubit interactions within this
engineered phononic environment.

On the benchmarking side, we defined candidate acceptance
targets---including single-hole control in each dot, $T_2^\ast$ and $T_2$
in the microsecond to tens-of-microseconds range for first-generation suspended devices,
relaxation-limited coherence as a longer-term target in high-purity isotopically enriched Ge,
enhanced $T_1$ within the PnC bandgap, and two-qubit coupling rates in the few-hundred-kHz
to MHz regime---that will test whether this platform can become competitive with existing
Si and Ge spin-qubit implementations operating at comparable temperatures. A particularly important benchmark is the joint measurement of
$T_1$, $T_2^*$, and $T_2$ at the same compensated operating point.
The leading transverse cavity interaction is tested through exchange
dynamics and its detuning dependence, with direct tunneling and thermal
occupation included. A longer $T_1$ or an avoided crossing alone does
not establish a coherent phonon-mediated gate.
Because the architecture is compatible with frequency-multiplexed RF readout and
with the integration of on-chip superconducting resonators,%
~\cite{hornibrook2014multiplexed,mi2018cavity,landig2019coherent}
it naturally supports future extensions toward hybrid spin--phonon--photon systems
and scaled arrays of Ge qubits.

Looking ahead, the two-qubit module developed here can serve as a foundational
building block for larger-scale quantum processors and quantum sensors based on
Ge. Systematic exploration of alternative PnC geometries, cavity designs, and
coupling schemes—guided by the experimental benchmarks proposed in
Sec.~\ref{sec:benchmarks}—will clarify the advantages and trade-offs of
phonon-mediated interactions relative to more conventional exchange or
capacitive coupling mechanisms.%
~\cite{maier2013relaxation,mi2018cavity,landig2019coherent}
The approach of jointly designing materials, phonons, and qubits while
specifying fabrication constraints and measurable performance targets may
also guide other solid-state platforms that use engineered phonons.

The proposed module provides a conditional route to investigate
phonon-mediated spin interactions. The new orbital sensitivity and
thermal-channel calculations identify failures of particular assumed
parameter combinations and pulse sequences. Equal-coupling exchange
within the single-excitation subspace can survive common Stark noise
without preserving arbitrary-input two-qubit coherence. Consequently,
exchange observation and full-gate validation are separate objectives. A validated $Q_m$ sweep over
$0$--$10^\circ$ sidewall taper, a self-consistent gate-potential model,
and measurements of the released strain and thermal mode occupation
remain essential before claiming warm-operation performance.

\appendix
\section{Thermal number-noise channel for specified exchange sequences}
\label{app:thermal_channel}
The dispersive frequency shift per oscillator quantum is $2\chi_i$.
Matching the mean dressed frequencies does not remove the fluctuating
differential shift
\begin{equation}
\begin{aligned}
 \delta\omega_{12}(t)&=2(\chi_1-\chi_2)[n(t)-\bar n],\\
 \sigma_{\delta\omega}&=2|\chi_1-\chi_2|\sqrt{\bar n(\bar n+1)}.
\end{aligned}
 \label{eq:differential_thermal_stark}
\end{equation}
For a single dispersively coupled spin and an equilibrium damped mode,
the long-time unrefocused thermal dephasing rate is~\cite{clerk2007thermal}
\begin{equation}
 \Gamma_{\phi,\rm th}=\frac{\kappa}{2}\operatorname{Re}
 \left[\sqrt{\left(1+\frac{2i\chi_i}{\kappa}\right)^2
               +\frac{8i\chi_i\bar n}{\kappa}}-1\right].
 \label{eq:thermal_phase_exact_dispersive}
\end{equation}
For the $5$ MHz, $100$ MHz-detuned example,
$\chi_i/2\pi=0.25$ MHz and this inverse rate is approximately
$0.445~\mu$s at $1$ K and $0.177~\mu$s at $4$ K.
These are asymptotic reduced-model dephasing scales, not Hahn-echo
times or measured device coherence. They show why a population lifetime
alone is insufficient to assess this operating point.

We also evaluate two specified ideal exchange sequences in the same
second-order model: an uninterrupted pulse of duration
$t_g=\pi/(4|J_c|)$, and a pulse with a simultaneous $X_1X_2$ operation
at its midpoint. The known final $X_1X_2$ is undone or included in the
target frame. In this frame the midpoint pulse leaves transverse
exchange unchanged and reverses the Stark terms. The durations and errors of the ideal single-qubit refocusing and
frame-correction pulses are neglected; the exchange duration is retained. Starting with a factorized thermal
mode, its number-diagonal, two-spin density operators obey
\begin{align}
 \dot\rho_n={}&-i[H_n/\hbar,\rho_n]
 +\kappa(\bar n+1)[(n+1)\rho_{n+1}-n\rho_n]\nonumber\\
 &+\kappa\bar n[n\rho_{n-1}-(n+1)\rho_n],\label{eq:number_blocks}\\
 \frac{H_n}{\hbar}={}&J_c(\sigma_1^+\sigma_2^-+\mathrm{h.c.})
 +y(t)(n-\bar n)\sum_i\chi_i\sigma_i^z,\label{eq:toggled_disp_model}
\end{align}
where $y=1$ for the plain sequence and changes from $+1$ to $-1$ at
the echo midpoint. This calculation isolates thermal number-induced
phase noise. It omits cavity-induced population transitions, intrinsic
spin noise, direct interaction, pulse imperfections, and higher-order
Jaynes--Cummings terms; it is not a complete device gate simulation.
Tracing the mode gives a qubit channel $\mathcal E$. For the specified
target $U=\exp[-i\pi(\sigma_1^+\sigma_2^-+\mathrm{h.c.})/4]$, its
average gate fidelity is calculated from the column-vectorized
superoperators as
\begin{equation}
 F_{\rm av}=\frac{\operatorname{Re}\operatorname{Tr}
 (S_U^\dagger S_{\mathcal E})+d}{d(d+1)},\qquad d=4.
 \label{eq:channel_average_fidelity}
\end{equation}
We report both the mean-frequency-compensated result and a result
allowing numerically optimized, fixed local $Z$ corrections. These
are deterministic phase calibrations, not feedback conditioned on
oscillator number.

The example uses $f_c=6$ GHz, $Q_m=18000$, and either
$g_1/2\pi=g_2/2\pi=5$ MHz or $g_1/2\pi=5$ MHz and
$g_2/2\pi=4.5$ MHz. For unequal couplings, the mean bare detuning remains
$100$ MHz and a small differential detuning is solved so that
$\omega_1+(2\bar n+1)\chi_1=\omega_2+(2\bar n+1)\chi_2$.
The resulting detunings divided by $2\pi$ are
$(99.831,100.169)$ MHz at $1$ K and $(99.296,100.704)$ MHz at $4$ K.
The associated differential Stark rms values are $0.334$ and $1.408$
MHz, respectively, whereas $2J_c/2\pi\simeq0.450$ MHz.

\begin{table}[t]
\centering
\caption{Conditional channel diagnostics for the specified ideal
sequences. $F_{\rm av}$ is the four-dimensional average gate fidelity
relative to the ideal exchange entangler, allowing calibrated fixed
local $Z$ corrections. $F_B$ is the fidelity of the ideal Bell-like
state produced from $|10\rangle$, before this additional phase
calibration. These are reduced-model diagnostics, not expected or
measured device fidelities; omitted effects are stated in the text.}
\label{tab:thermal_channel_diagnostic}
\begin{tabular}{cccrr}
\hline
$T$ (K)& $g_2/g_1$ & Sequence & $F_{\rm av}$ & $F_B$\\
\hline
1 & 1.0 & Plain & 0.585 & 1.000\\
1 & 1.0 & Echo  & 0.659 & 1.000\\
1 & 0.9 & Plain & 0.552 & 0.927\\
1 & 0.9 & Echo  & 0.640 & 0.970\\
4 & 1.0 & Plain & 0.508 & 1.000\\
4 & 1.0 & Echo  & 0.513 & 1.000\\
4 & 0.9 & Plain & 0.376 & 0.608\\
4 & 0.9 & Echo  & 0.427 & 0.750\\
\hline
\end{tabular}
\end{table}

For equal $\chi_i$, the single-excitation subspace has zero common
Stark shift, so its ideal Bell-state diagnostic is protected in this
restricted model even when arbitrary-input gate coherence is poor.
Unequal $\chi_i$ removes that protection. The specified midpoint echo
does not bring the full-channel diagnostic near $90\%$ at either
temperature. This result does not exclude other pulse sequences,
cooling, encoding, or operating points. It does mean that the present
passive thermal example does not justify a $90$--$95\%$ gate-performance
expectation. Any such target requires a separately specified protocol
and a complete channel calculation with measured parameters. The
calculation files include oscillator-cutoff convergence and numerical
trace-preservation, complete-positivity, and equal-coupling symmetry
checks.

\section{Electrostatic spreading in an ideal strip-gate limit}
\label{app:strip_gate_diagnostic}

An exactly specified electrostatic limit illustrates why a lithographic
linewidth does not set the barrier width in the well. Consider a uniform,
charge-free dielectric occupying $z>0$, invariant along $y$. An infinitely
long strip of width $w$, centered at $x=0$ in the plane $z=0$, has potential
$V_g$; the rest of that plane, including gaps, is grounded. This is a
separate Dirichlet boundary-value problem. It omits finite gate lengths,
the gate dielectric and material interfaces, fixed/mobile charge,
reservoirs, and the released strain profile. Its sampling depth $z=40$ nm
matches the nominal semiconductor surface-to-well-center distance, before
dielectric thickness; it is not a representation of the actual gate stack.

The Poisson kernel for Laplace's equation gives
\begin{align}
 \phi(x,z)&=\frac{1}{\pi}\int_{-\infty}^{\infty}
 \frac{z\phi(x',0)}{(x-x')^2+z^2}\,dx',\nonumber\\
 &=\frac{V_g}{\pi}\left[
 \arctan\frac{x+w/2}{z}-\arctan\frac{x-w/2}{z}\right].
 \label{eq:strip_gate_potential}
\end{align}
For this isolated strip, solving $\phi(x_{1/2},z)=\phi(0,z)/2$ gives
\begin{equation}
 W_{1/2}=2x_{1/2}=\sqrt{w^2+4z^2},\qquad
 \frac{\phi(0,z)}{V_g}=\frac{2}{\pi}\arctan\frac{w}{2z}.
 \label{eq:strip_gate_width}
\end{equation}
At $z=40$ nm, $w=20$, $35$, and $50$ nm give full widths at half maximum
of $82.46$, $87.32$, and $94.34$ nm, respectively. At the two reference
positions $x=\pm50$ nm, the potential increment is $40.3\%$--$46.7\%$
of its central value. The 15 nm FWHM cosine barrier in the orbital
calculation therefore cannot be identified with an isolated strip's
response in this limit. This is not a universal minimum feature width:
opposite-sign neighboring gates and different boundaries can reshape
the potential.

To show that distinction explicitly, a second algebraic example uses
three 20 nm strips centered at $x=-50,0,+50$ nm, with the same grounded
remaining plane and $z=40$ nm. Voltage increments of approximately
$(-0.190,+0.538,-0.190)$ V impose
$\delta\phi(-50,40)=\delta\phi(50,40)=0$ and
$\delta\phi(0,40)=60$ mV. For holes, $\delta U=e\delta\phi$.
The central positive lobe has a 53.77 nm FWHM, but the energy gradients
at $x=\pm50$ nm are $\mp0.750$ meV/nm. Thus these two fixed-point
potential constraints do not preserve the original potential minima
or orbital curvatures. The voltages are solutions of this idealized
boundary-value problem, not allowed biases or calibrated lever arms
for the proposed device.

\begin{figure*}[t]
 \centering
 \includegraphics[width=0.96\textwidth]{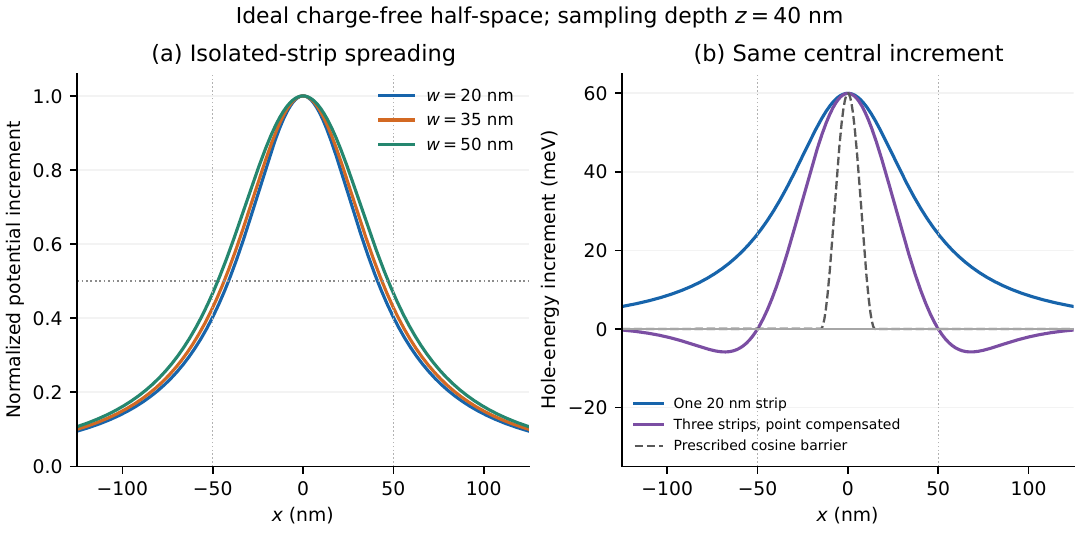}
 \caption{Analytical electrostatic diagnostic in a uniform, charge-free
 half-space sampled at $z=40$ nm. (a) Normalized responses of isolated
 infinite strips. Dotted vertical lines mark $x=\pm50$ nm.
 (b) Hole-energy increments normalized to a 60 meV center value for a
 single 20 nm strip, the three-strip point-compensated construction,
 and the prescribed cosine barrier used in the orbital model.
 These are increments, not complete confinement potentials.
 Opposite-sign strips sharpen the response while changing its gradients.
 No charge occupancy, wavefunction, spin-frequency, voltage-range, or
 spin--phonon-coupling validation is inferred from these curves.}
 \label{fig:strip_gate_diagnostic}
\end{figure*}

The script and profile arrays supply the analytical geometry, voltages,
formula checks, and resulting gradients. This diagnostic does not supply
the physical-barrier-voltage wavefunctions or finite joint tuning
trajectory requested by the referee. Those require the actual device
boundary conditions, self-consistent potential and multiband states,
followed by charge and both spin--phonon coupling tests on the same path.

\section{Requirements for validation of one candidate device}
\label{app:common_validation}
\label{subsec:common_device_validation}

Closing the mechanical and control questions requires one documented
candidate device shared by the calculations. Selecting its missing
features is remaining design work. The 150/20/30 nm semiconductor core,
nominal 1 $\mu$m pitch, 300 nm top radius and 100 nm dot separation are
reference inputs. They do not specify the retained etch stop $E$, defect
pattern, finite shield, supports, gate footprints or coating loads.
These choices, material tensors and stress references must be recorded
before a result can be assigned to the proposed module. The assumed
MHz couplings and $Q_m=1.8\times10^4$ are comparison values, not boundary
conditions or outputs of that calculation.

The mechanical model must follow support removal and cooling, with
lattice mismatch, deposited-film stress and thermal contraction treated
consistently. It must check stable equilibrium and report the deflection
and full strain tensor at both dot locations. The resulting geometry
and prestress then enter the acoustic calculation. Unit-cell bands
test the intended stopband; a finite structure with its actual anchors
must separately establish a localized baseline mode and radiation loss.
The orbital and spin calculations must use that same released well,
gate geometry and normalized strain field, rather than combining
independently favorable examples.

At fixed top-mask radius, the finite-device calculation must track the
baseline mode through $\theta=0,2,4,6,8,10^\circ$ and report frequency,
$Q_{\rm rad}$, localization and numerical uncertainty. Nonmonotonic
behavior, hybridization or disappearance of the localized mode must
be retained. At least three mesh levels and independent changes of
substrate extent and PML settings should establish convergence; the
remaining uncertainty must be smaller than any claimed taper effect.
Intrinsic and loading loss enter a separate total-$Q$ budget. The
package specifies the required inputs and outputs, but supplies no
finite-device strain map, localized-mode solution or $Q(\theta)$ result.
The electronic calculation must use the complete electrode masks and
connections, dielectric and fixed-charge inputs, reservoir boundary
conditions, usable voltage limits, and the released multiband well profile.
For the nominal and offset cavity frequencies, a finite compensated
trajectory must report all gate voltages, both dressed qubit frequencies,
competing charge-sector energy costs, carrier temperature, tunnel and
direct-coupling bounds, and both spin--phonon couplings at every accepted
point. Potentials and wavefunctions at several physical barrier-gate
voltages must come from that same geometry and bias sequence. The
prescribed barrier energies and the ideal strip-gate calculation in
Appendix~\ref{app:strip_gate_diagnostic} do not supply those device results.

The proposed numerical evidence can be obtained before fabricating the
complete module, provided its physical assumptions and uncertainty are
stated. Process compatibility and experimental coherence still require
separate measurements. Completing these calculations is necessary before
claiming the corresponding device feasibility; a future experimental
protocol does not supply the missing $T_2^*$ or $T_2$ measurements.

\paragraph{Preliminary feasibility screens and evidence status.}
The known illustrative $\pm150$ MHz offset would correspond to
$|\delta g|=0.05359$ at $B=0.20$ T if compensated entirely by a
local Zeeman shift. For a single competing charge sector with unit
multiplicity, $\epsilon_{\rm ch}=0.01$ requires
$\Delta E\ge0.397$ meV at $1$ K and $\Delta E\ge1.587$ meV
at $4$ K in the simple equilibrium-weight approximation. These are
screening thresholds, not calculated properties of the proposed gates.
The necessary evidence is explicitly separated as follows:
\begin{itemize}
\item \emph{Completed:} core-only laminate strain/curvature estimates,
frustum-angle geometry, prescribed-potential orbital and tunneling
calculations, ideal electrostatic spreading, and reduced thermal-channel
calculations.
\item \emph{Not completed:} full-stack support-constrained release
simulation and dot-local strain; prestressed cavity frequency and the
converged $Q_{\rm rad}(\theta)$ sweep; a physical-gate $(1,1)$ trajectory
with both $g_i$ retained; and experimental $T_2^*,T_2$ data.
\end{itemize}
The latter results are prerequisites to claiming experimentally
qualified fabrication tolerances or high-fidelity warm operation.

\section*{Data and code availability}
The code and numerical data supporting this study are supplied in the
accompanying Supplementary Code and Data archive. It contains the
prescribed-potential orbital calculations and sensitivities, core-only
laminate and taper calculations, the explicitly idealized five-strip
trajectory and edge-damped acoustic-chain calculations, thermal-channel arrays, figure data,
and numerical convergence records, together with instructions for
reproducing the reported results. No experimental data are reported.
The archive distinguishes these completed calculations from the
unperformed finite-device release, electrostatic tuning, and
$Q_{\rm rad}(\theta)$ studies specified in
Appendix~\ref{app:common_validation}; it contains no datasets for those
unperformed studies.

\section*{Acknowledgments}
This work was supported in part by NSF OISE 1743790, NSF PHYS 2117774, NSF OIA 2427805, NSF PHYS 2310027, NSF OIA 2437416, DOE DE-SC0024519, DE-SC0004768, and a research center supported by the State of South Dakota. 

\bibliographystyle{apsrev4-2}
\bibliography{refs}

\end{document}